\documentclass[twocolumn]{aastex62}

\newcommand\afe{[$\alpha$/Fe]}

\newcommand\bafe{[Ba/Fe]}
\newcommand\bafer{[Ba/Fe]$_r$}
\newcommand\bah{[Ba/H]}
\newcommand\feh{[Fe/H]}
\newcommand\mgfe{[Mg/Fe]}
\newcommand\baeu{[Ba/Eu]}
\newcommand\eufe{[Eu/Fe]}
\newcommand\teff{$T_{\rm{eff}}$}
\newcommand\logg{$\log g$}
\newcommand\loggf{$\log gf$}
\newcommand\msun{M$_\odot$}
\newcommand\mstar{$M_\ast$}
\newcommand\rproc{$r$-process}
\newcommand\sproc{$s$-process}

\graphicspath{{./}}

\received{July 16, 2018}
\revised{September 12, 2018}
\accepted{XXX}
\submitjournal{ApJ}

\shorttitle{Early Chemical Evolution of Dwarf Galaxies}
\shortauthors{Duggan et al.}

\begin{document}

\title{Neutron Star Mergers Are the Dominant Source of the r-process in the Early Evolution of Dwarf Galaxies}

\correspondingauthor{Gina E. Duggan}
\email{gduggan@astro.caltech.edu}

\author[0000-0002-9256-6735]{Gina E. Duggan}
\affil{California Institute of Technology \\ 
1200 E. California Boulevard, MC 249-17 \\ 
Pasadena, CA 91125, USA}

\author[0000-0001-6196-5162]{Evan N. Kirby}
\affiliation{California Institute of Technology \\ 
1200 E. California Boulevard, MC 249-17 \\ 
Pasadena, CA 91125, USA}

\author{Serge M. Andrievsky}
\affiliation{Astronomical Observatory\\
Odessa National University\\ 
Shevchenko Park, 65014 Odessa, Ukraine}
\affiliation{GEPI, Observatoire de Paris, PSL, Research University, \\
CNRS, Univ Paris Diderot, Sorbonne Paris Cit\'e \\ 
Place Jules Janssen, 92195, Meudon, France}

\author[0000-0002-4058-8780]{Sergey A. Korotin}
\affiliation{Crimean Astrophysical Observatory \\ 
Nauchny, 298409, Crimea}

\begin{abstract}

There are many candidate sites of the \rproc: core-collapse supernovae (including rare magnetorotational core-collapse supernovae), neutron star mergers, and neutron star/black hole mergers. The chemical enrichment of galaxies---specifically dwarf galaxies---helps distinguish between these sources based on the continual build-up of \rproc\ elements. This technique can distinguish between the \rproc\ candidate sites by the clearest observational difference---how quickly these events occur after the stars are created. The existence of several nearby dwarf galaxies allows us to measure robust chemical abundances for galaxies with different star formation histories. Dwarf galaxies are especially useful because simple chemical evolution models can be used to determine the sources of \rproc\ material. We have measured the \rproc\ element barium with Keck/DEIMOS medium-resolution spectroscopy. We present the largest sample of barium abundances (almost 250 stars) in dwarf galaxies ever assembled. We measure \bafe\ as a function of \feh\ in this sample and compare with existing \afe\ measurements. We have found that a large contribution of barium needs to occur at more delayed timescales than core-collapse supernovae in order to explain our observed abundances, namely the significantly more positive trend of the \rproc\ component of \bafe\ vs. \feh\ seen for $\text{\feh}\lesssim-1.6$ when compared to the \mgfe\ vs. \feh\ trend. We conclude that neutron star mergers are the most likely source of \rproc\ enrichment in dwarf galaxies at early times.

\end{abstract}


\section{Introduction}
\label{sec:ba}

The premise of galactic archeology is that stars form out of a galaxy's gas and adopt the chemical composition of the gas when they were born. At the same time, galactic inflows, outflows, and stellar ejecta dilute or enhance the chemical composition in a galaxy. We can not watch these processes happen in real time in an individual galaxy, so instead we use galactic archeology to learn how a given galaxy has evolved. Galactic archeology looks back in time by utilizing the fact that long-lived stars retain the memory of the chemical composition of the galaxy when they were born. We trace a variety of elements in stars alive today to learn about the timescales and origins of chemical enrichment in dwarf galaxies. 

We will discuss three different groups of elements observable in stars. These elements were chosen because each group traces different stellar events. The first group is the $\alpha$ elements (e.g., Mg). These elements are created and ejected in core-collapse supernovae (CCSNe) with a short delay time ($4-25$\,Myr). The delay time is defined as the time between the star being born and the star ejecting enriched material into the ISM. The second group is iron. Iron is produced in large amounts by SNe~Ia with a relatively long delay time \citep[$0.04-14$\,Gyr,][]{Maoz17} described by a delay time distribution. In addition, there are moderate amounts of iron that are generated in each CCSNe. Finally, elements heavier than iron are formed through neutron-capture processes. Determining the dominant origin of neutron-capture elements is still an active area of study and is the focus of this paper. Because the origin is unknown, the delay time is also unknown.

The ratio of $\alpha$ elements to iron (\afe) is commonly used as a chemical clock \citep{Tinsley80}. If we know the exact amount of each element released during each process (also referred to as the yield), [$\alpha$/Fe] can tell us the ratio of CCSNe to SNe~Ia as a function of time. Starting from a star formation history (SFH), one can convert time into an iron abundance by assuming an initial mass function (IMF) and supernovae iron yields. As a stellar population ages, the rate of CCSNe changes compared to SNe~Ia because of their different lifetimes. The \afe\ starts out high at low \feh\ (or early times) because CCSNe quickly eject a large amount of $\alpha$ elements with small amounts of iron \citep{Nomoto06}. Then \afe\ dramatically declines as time passes and \feh\ increases, because SNe~Ia start to explode and eject large amounts of iron \citep{Iwamoto99}. The plot of \afe\ vs. \feh\ has been used in many studies of galactic evolution, typically by using \mgfe\ as an indicator of the total \afe\ \citep[e.g.,][]{Gilmore91,Shetrone01,Venn04,Kirby11,Bensby14}.

Using neutron-capture elements as a chemical clock is less common, but may be the key to distinguish the dominant origin of neutron-capture elements. This paper uses barium as a tracer of all neutron-capture elements, because barium is arguably the easiest neutron-capture element to measure due to its several strong absorption lines available in the optical. The \bafe\ indicates what levels of barium are being ejected into the ISM compared to SNe~Ia throughout time. Combining the \bafe\ vs. \feh\ \lq chemical clock' with abundances of another neutron-capture element (e.g., europium), clarifies the origin of neutron-capture elements even further. The ratio of these two different neutron-capture elements (\baeu) tells us the percentage of all neutron-capture elements produced by the two different neutron-capture processes: the $s$- and the \rproc. 

The slow neutron-capture process (\sproc) and the rapid neutron-capture process (\rproc) occur in very different physical scenarios and originate in very different astrophysical origins. The \sproc\ occurs in episodes lasting from $10^2-10^4$\,yrs. The \sproc\ has a very long time between each free neutron being captured ($\tau_{\text{n}}\sim10-10^4$\,yr), and this timescale is much longer than the timescale for that neutron to $\beta$-decay into a proton. We know that for heavy neutron-capture elements---such as barium---the \sproc\ is produced by AGB stars, with trace amounts possibly produced in massive stars \citep[e.g.,][]{Karakas14}.

The \rproc\ occurs in a single event that lasts seconds and an isotope can capture many neutrons ($\tau_{\text{n}}\sim10^{-2}-10$\,sec) before those neutrons $\beta$-decay into a proton. Unlike the \sproc, there is little consensus on the astrophysical origin of the \rproc, but the origin has been isolated to various explosions or mergers. We will now use the results of theoretical models and observational constraints to limit our search for the dominant \rproc\ origin.

\subsection{Narrowing Down the Search for the Dominant R-Process Origin}
\label{sec:potential_sources}

Arguably the largest gap in our knowledge of stellar nucleosynthesis is the origin of heavy elements that are produced by the \rproc. Specifically, CCSNe, magnetorotational supernovae (MRSNe), common envelope jets SNe, binary neutron star mergers (NSMs), and neutron star/black hole mergers (NS+BHs) are currently being considered \citep[e.g.,][]{Arnould07,Thielemann11,Papish15,Liccardo18}. 

In a CCSN it was thought that neutrino winds drive neutrons and protons from the surface of the proto-neutron star (located at the core), resulting in a large neutron flux in the middle region of the explosion, which might be a site of \rproc\ \citep{Woosley05}. However, recent simulations have been unable to generate \rproc{} elements up to or beyond barium except in extreme cases \citep[e.g.,][]{Wanajo13}. This limitation on the average \rproc\ yields of CCSNe provided by simulations paired with estimates of the yield required by observations has convincingly eliminated \lq typical' CCSNe as the dominant source of the \rproc\ \citep[e.g.,][]{Macias16}. 

However, rare types of CCSNe that produce copious amounts of \rproc\ material are still being considered as a potential source of \rproc. MRSNe are supernovae that start out with high magnetic fields (B$\sim 10^{12-13}$\,G) that cause jet-like explosions, which may produce \rproc\ elements. MRSNe are possibly $0.1\%-1\%$ of all CCSNe and could produce enough \rproc\ material to account for the observed levels of \rproc\ enrichment found in the solar system. Simulations have been able to reproduce the full \rproc\ pattern seen in the solar system with only minor discrepancies \citep{Mosta17,Nishimura15,Nishimura17}. There is currently no proof that MRSNe occur, although it is the most popular explanation for hydrogen-poor superluminous supernovae \citep[SLSNe~I, e.g.,][]{Kasen10,DeCia18}. This lack of observational constraints means that there is also no proof that MRSNe actually produce large amounts of \rproc\ material.

Another rare type of SN is a common envelope jets SN \citep{Papish15,Soker17}.  In this scenario, the primary star in a binary system evolves into a neutron star.  When the secondary becomes a red supergiant, it subsumes the neutron star, which spirals into the giant.  If the neutron star makes it to the core, it can launch jets and form an accretion disk, possibly including $r$-process nucleosynthesis \citep{Grichener18}.  This type of SN would have a delay time slightly longer than a CCSN but shorter than a NSM\@.

NSMs have the most complete theoretical models and easily produce a large amount of \rproc\ material. Simulations have predicted that NSMs are able to recreate the full \rproc\ pattern (including barium), and have shown that these results are insensitive to the detailed choices of the simulation \citep{Cote17}. NSMs now have an observational constraint in the form of the first observed neutron star merger \citep[GW170817,][]{Abbott17}. The discovery of the electro-magnetic counterpart was originally announced by \citet{Coulter17}. Photometric and spectral follow-up enabled \rproc\ yields to be determined \citep{Tanvir17,Troja17,Evans17}. This is the first observational constraint of an \rproc\ yield coming from a known astrophysical origin. This NSM yield is high enough for all \rproc\ elements to be produced by NSMs. However, assuming the yield from GW170817 is an accurate average of all NSMs could be inaccurate. Quantitative predictions for the rates, time delays, and yields are still active areas of study \citep[e.g.,][]{Radice16}. 

Some NS+BHs likely reproduce the full \rproc\ pattern as well, but restrictions (e.g., on the black hole mass and spin) are required to eject \rproc\ material from a NS+BHs \citep[e.g.,][]{Lattimer74,Shibata11,Foucart15,Barack18}. The subset of NS+BHs that are a source of \rproc\ are expected to be much rarer than the NSMs. Therefore, the contribution of neutron star--neutron star mergers dominates over the contribution of neutron star--black hole mergers. 

We have narrowed down our search for the dominant source of \rproc\ enrichment in galaxies to a rare form of core-collapse supernovae (i.e., MRSNe) or NSMs. The simulations of these two candidate sites are so poorly constrained that distinguishing between MRSNe and NSMs by comparing detailed abundance patterns is extremely challenging \citep[e.g.,][]{Ji18}. The clearest way to definitively distinguish between MRSNe and NSMs is by their different timescales.

\subsection{Distinguishing Between Dominant R-Process Candidates Based on Timescales for the First Time}

The chemical enrichment of galaxies---specifically dwarf galaxies---enables us to distinguish between these sources based on their timescales. This is possible because we observe the continual build-up of \rproc\ elements, and this type of study is an essential counterpart to the characterization of individual events. Specifically, we are sensitive to the enrichment timescale, which is the key distinguishing characteristic between MRSNe and NSMs. Robust chemical abundance trends can be measured for nearby dwarf galaxies, because their nearness and intact stellar populations allow us to average the abundances of many individually resolvable stars. Determining the sources of chemical enrichment from our observed abundance trends in dwarf galaxies is possible, because (1) their small masses make them very sensitive to feedback mechanisms, (2) the lack of major mergers helped preserve their stellar populations, and (3) their small sizes result in nearly instantaneous mixing compared to the chemical enrichment timescale \citep{Escala18}. All of these properties make dwarf galaxies the perfect test sites to observe the simplest form of galactic chemical evolution. The existence of several nearby dwarf galaxies provides additional diagnostic power, because we can measure the chemical enrichment for galaxies with different star formation histories. The lessons learned in dwarf galaxies can then be applied to larger, more complex galaxies.  

The usefulness of dwarf galaxies for determining the characteristics of \rproc\ enrichment can be seen in \citeauthor{Ji16}'s \citeyear{Ji16} study of the ultra-faint dwarf (UFD) galaxy Reticulum~II. Several stars in Reticulum~II have very high levels of enrichment in barium and europium. No other similarly enhanced stars were found in the other nine UFD galaxies considered. This indicates that a rare event occurred that dramatically increased the neutron-capture enrichment in Reticulum~II. The \baeu\ confirms that this enrichment was created by the \rproc\@. \citet{Ji16} calculated that this high enrichment could be caused by a single event in the small UFD galaxy. This rules out a typical CCSN, because CCSNe are so frequent that we would see this effect in many UFD galaxies \citep[also see][]{Beniamini16b,Beniamini18}. Both MRSNe and NSMs are predicted to be rare and produce a large amount of \rproc\ enrichment. As we have mentioned before, the main observable difference between MRSNe and NSMs is their timescales. The short SFH of Reticulum~II challenges whether it is possible to have a NSM occur while stars are still forming. However, because we are discussing a single rare event, it is possible an unusually quick NSM occurred in Reticulum~II. Therefore we need to see enrichment occur in larger mass dwarf galaxies, so that many of these rare prolific \rproc\ events have occurred and we have some statistical certainty if we need an event with a short (MRSNe) or long (NSMs) timescale.

In order to break this degeneracy between NSMs and MRSNe, we need a large sample of stars in many moderately-sized dwarf galaxies. We have measured barium with DEIMOS medium-resolution spectroscopy \citep{Faber03}, and will present the largest sample of barium abundances (almost 250 stars) in dwarf galaxies ever assembled. 

\subsection{Abundance Measurements from Medium-Resolution Multi-Object Spectroscopy}

A large number of stars is needed to distinguish the chemical trend of a stellar population from star-to-star variations. We were able to obtain a large sample of stars, because we used multi-object, medium-resolution spectroscopy (MRS, R $\approx 5,000$). Typically barium is measured using high-resolution spectroscopy (HRS, R $>20,000$) on single-slit spectrographs. Traditionally, precise chemical abundance measurements required equivalent width measurements of absorption lines in HRS, and because of this the Dwarf Abundances and Radial velocities Team \citep[DART;][]{Tolstoy06} invested the necessary observing time to obtain HRS for tens of individual stars in Sculptor and Fornax \citep[e.g.,][]{Lemasle14,Starkenburg13}. To obtain a large sample of stars with moderate observing time, we used MRS which enables tens of member stars to be observed simultaneously. 

The main weakness of MRS is increased line blending. The blending that occurs in MRS causes strong sky lines to contaminate a larger range of wavelengths, the continuum to be obscured, and the apparent weakening of absorption lines. Since blending prevents the continuum from being measured in the gaps between absorption/emission lines, the continuum is iteratively fitted while the abundances are measured using the synthetic spectra. 

We overcome these weaknesses to take advantage of higher S/N per pixel or fainter limiting magnitude that can be achieved with a given amount of observation time ($V \lesssim 20$\,mag compared to 18\,mag for HRS), which increases the sample of observable red giant branch stars. Wider spectral coverage increases the number of absorption lines observed per chemical element, which is especially important for elements with few clear absorption lines (e.g., neutron-capture elements). For barium, we use five different absorption lines in our measurements. 

Both HRS and MRS require a stellar atmosphere model and stellar line analysis---either to measure abundances from equivalent widths or to generate synthetic spectra. Common simplifications for both methods are to use a one-dimensional (1D) stellar atmosphere model and to assume local thermodynamic equilibrium (LTE) throughout the star. Correcting for 3D and non-LTE effects could systematically shift barium abundance measurements by 0.1--0.3\,dex (see Section~\ref{sec:nonLTE}), similar to the statistical uncertainties of our measurements. Unfortunately, both of these corrections are very computationally intensive and are beyond the scope of this project. 

Determining chemical abundances from MRS has returned uncertainties as low as 0.1\,dex for iron and $\alpha$~elements in dwarf galaxies \citep{Kirby10}. This paper will demonstrate that we can also achieve similar uncertainties for barium abundances in dwarf galaxies. Therefore, we are able to use a multi-object spectrographs MRS without sacrificing accuracy to obtain barium abundances for a large sample of stars.

\section{Observations}

We observed a diverse sample of dwarf galaxies to probe what \bafe\ measurements can tell us about the chemical enrichment mechanisms and SFHs in different galaxies. Our sample includes five classical dwarf spheroidal galaxies: Fornax, Sculptor, Sextans, Draco, and Ursa Minor. 
These galaxies span a variety of masses (M$_\ast\sim 10^5-10^7$\,\msun) and durations of star formation ($\approx 1$--$11$\,Gyr of star formation, \citealt{Weisz14}). In each of these galaxies we obtained MRS for individual red giant branch stars using DEIMOS on Keck II. 

Apart from these galaxies, we also observed red giant branch stars in globular clusters and the halo of the Milky Way to compare our \bafe\ measurements to those found in the literature. This comparison is used to estimate our systematic error (see Section \ref{sec:HRS_MRS}).

The locations and distances to all spectroscopic targets are listed in Table \ref{tab:targets}. The details of all observations contained in this paper are given in Table \ref{tab:obs}. This includes the name of the slitmask, number of slits, date, airmass, seeing, and exposure time of all observations. This table also includes references for star selection and membership verification. 

\begin{deluxetable}{lc@{~~}c@{~~}c@{~~}c}
\tablewidth{0pt}
\tablecolumns{5}
\tablecaption{Spectroscopic Targets\label{tab:targets}}
\tablehead{\colhead{Target} & \colhead{RA} & \colhead{Dec} & \colhead{D} & $(m-M)_0$\tablenotemark{a} \\
                            & \colhead{(J2000)} & \colhead{(J2000)} & \colhead{(kpc)} & \colhead{(mag)}}
\startdata
\cutinhead{Globular Clusters}
NGC 2419          & $07^{\mathrm{h}} 38^{\mathrm{m}} 09^{\mathrm{s}}$ & $+38\arcdeg 52\arcmin 55\arcsec$ &       \phn82.6 & 19.83 \\ 
NGC 4590   (M68)  & $12^{\mathrm{h}} 39^{\mathrm{m}} 28^{\mathrm{s}}$ & $-26\arcdeg 44\arcmin 39\arcsec$ &     \phn10.3 & 15.21 \\
NGC 6341 (M92)    & $17^{\mathrm{h}} 17^{\mathrm{m}} 07^{\mathrm{s}}$ & $+43\arcdeg 08\arcmin 11\arcsec$ &    \phn\phn8.3 & 14.65 \\ 
NGC 7078 (M15)    & $21^{\mathrm{h}} 29^{\mathrm{m}} 58^{\mathrm{s}}$ & $+12\arcdeg 10\arcmin 01\arcsec$ &       \phn10.4 & 15.39 \\ 
\cutinhead{Halo Field Stars}
BD $+$14 550  & $03^{\mathrm{h}} 18^{\mathrm{m}} 27^{\mathrm{s}}$ & $+15\arcdeg 10\arcmin 38\arcsec$ & & \\ 
BD $-$00 552   & $03^{\mathrm{h}} 28^{\mathrm{m}} 54^{\mathrm{s}}$ & $-00\arcdeg 25\arcmin 03\arcsec$ & & \\
BD +22 626  & $04^{\mathrm{h}} 04^{\mathrm{m}} 11^{\mathrm{s}}$ & $+23\arcdeg 24\arcmin 27\arcsec$ & & \\  
BD $-$13 942  & $04^{\mathrm{h}} 38^{\mathrm{m}} 56^{\mathrm{s}}$ & $-13\arcdeg 20\arcmin 48\arcsec$ & & \\
BD $-$14 1399  & $06^{\mathrm{h}} 18^{\mathrm{m}} 49^{\mathrm{s}}$ & $-14\arcdeg 50\arcmin 43\arcsec$ & & \\
BD $+$62 959  & $07^{\mathrm{h}} 54^{\mathrm{m}} 29^{\mathrm{s}}$ & $+62\arcdeg 08\arcmin 11\arcsec$ & & \\
BD $+$80 245  & $08^{\mathrm{h}} 11^{\mathrm{m}} 06^{\mathrm{s}}$ & $+79\arcdeg 54\arcmin 30\arcsec$ & & \\
BD $+$21 1969  & $09^{\mathrm{h}} 06^{\mathrm{m}} 43^{\mathrm{s}}$ & $+20\arcdeg 30\arcmin 36\arcsec$ & & \\
BD $-$20 2955  & $09^{\mathrm{h}} 36^{\mathrm{m}} 20^{\mathrm{s}}$ & $-20\arcdeg 53\arcmin 15\arcsec$ & & \\
BD $+$55 1362  & $10^{\mathrm{h}} 04^{\mathrm{m}} 43^{\mathrm{s}}$ & $+54\arcdeg 20\arcmin 43\arcsec$ & & \\
BD $+$54 1359  & $10^{\mathrm{h}} 14^{\mathrm{m}} 29^{\mathrm{s}}$ & $+53\arcdeg 33\arcmin 39\arcsec$ & & \\
BD $+$40 2408  & $11^{\mathrm{h}} 13^{\mathrm{m}} 55^{\mathrm{s}}$ & $+39\arcdeg 58\arcmin 40\arcsec$ & & \\
BD $-$04 3155  & $11^{\mathrm{h}} 51^{\mathrm{m}} 50^{\mathrm{s}}$ & $-05\arcdeg 45\arcmin 44\arcsec$ & & \\
BD $+$49 2098  & $11^{\mathrm{h}} 58^{\mathrm{m}} 00^{\mathrm{s}}$ & $+48\arcdeg 12\arcmin 12\arcsec$ & & \\
BD $+$09 2653  & $12^{\mathrm{h}} 40^{\mathrm{m}} 14^{\mathrm{s}}$ & $+08\arcdeg 31\arcmin 38\arcsec$ & & \\
\cutinhead{dSphs}
Sculptor          & $01^{\mathrm{h}} 00^{\mathrm{m}} 09^{\mathrm{s}}$ & $-33\arcdeg 42\arcmin 32\arcsec$ & \phn85\phd\phn & 19.67 \\ 
Fornax            & $02^{\mathrm{h}} 39^{\mathrm{m}} 59^{\mathrm{s}}$ & $-34\arcdeg 26\arcmin 57\arcsec$ &    139\phd\phn & 20.72 \\ 
Sextans           & $10^{\mathrm{h}} 13^{\mathrm{m}} 03^{\mathrm{s}}$ & $-01\arcdeg 36\arcmin 52\arcsec$ & \phn95\phd\phn & 19.90 \\ 
Ursa Minor        & $15^{\mathrm{h}} 09^{\mathrm{m}} 11^{\mathrm{s}}$ & $+67\arcdeg 12\arcmin 52\arcsec$ & \phn69\phd\phn & 19.18 \\ 
Draco             & $17^{\mathrm{h}} 20^{\mathrm{m}} 19^{\mathrm{s}}$ & $+57\arcdeg 54\arcmin 48\arcsec$ & \phn92\phd\phn & 19.84  
\enddata
\tablenotetext{a}{Extinction corrected distance modulus.}
\tablerefs{See \citet[][2010 edition, \url{http://www.physics.mcmaster.ca/~harris/mwgc.dat}]{Harris96} and references therein for the coordinates and distances for the globular clusters. 
Halo field stars coordinates are from \citet{Fulbright00}.
The remaining dSph coordinates are adopted from \citet{Mateo98}, and the distances are adopted from the following sources: Sculptor, \citet{Pietrzynski08}; Fornax, \citet{Rizzi07}; Sextans, \citet{Lee03}; Ursa Minor, \citet{Mighell99}; Draco, \citet{Bellazzini02}.}
\end{deluxetable}

\begin{deluxetable*}{lllcr@{ }c@{ }lccc}
\tablenum{2}
\tablewidth{0pt}
\tablecolumns{10}
\tablecaption{DEIMOS Observations\label{tab:obs}}
\tablehead{\colhead{Object} & \colhead{Slitmask} & \colhead{\phantom{\tablenotemark{a}}Reference\tablenotemark{a}} & \colhead{\# targets} & \multicolumn{3}{c}{Date} & \colhead{Airmass} & \colhead{Seeing} & \colhead{Exposures} }
\startdata
\cutinhead{Globular Clusters}
NGC~2419\tablenotemark{b} &  n2419b & \citet{Kirby16}  &      112   & 2012 & Mar & 19 & 1.07 & $0\farcs74$    & 3 $\times$ 900~s \\
NGC 4590 (M68)\tablenotemark{b} & n4590a & \citet{Kirby16} & \phn96 & 2014 & Feb & 2 & 1.60 & $0\farcs80$ & 1200~s, 937~s\\
NGC 6341 (M92)              & 6341l1 & \citet{Kirby16} & 177 & 2017 & Mar & 28  & 1.20 &    $0\farcs71$     & 6 $\times$ 1800~s, 2000~s \\
NGC 7078 (M15)\tablenotemark{b}              & n7078d & \citet{Kirby16}   &       164 & 2011 & Jul & 29 & 1.03 & $1\farcs10$    & 3 $\times$ 600~s \\*
                 & n7078e & \citet{Kirby16}   &       167 & 2011 & Jul & 30 & 1.05 & $0\farcs86$    & 3 $\times$ 900~s \\*
\cutinhead{Halo Field Stars}
BD $+$14 550     & LVMslits & \citet{Fulbright00} & \phn\phn1 & 2016 & Dec & 30 & 1.18 & $> 1\farcs5$ & 300~s \\*
BD $-$00 552     & LVMslits & \citet{Fulbright00} & \phn\phn1 & 2016 & Dec & 30 & 1.14 & $> 1\farcs5$ & 600~s \\* 
BD +22 626     & LVMslits & \citet{Fulbright00} & \phn\phn1 & 2016 & Dec & 30 & 1.05 & $> 1\farcs5$ & 600~s \\* 
BD $-$13 942     & LVMslits & \citet{Fulbright00} & \phn\phn1 & 2016 & Dec & 30 & 1.20 & $> 1\farcs5$ & 300~s \\*
BD $-$14 1399     & LVMslits & \citet{Fulbright00} & \phn\phn1 & 2016 & Dec & 30 & 1.26 & $> 1\farcs5$ & 2 $\times$ 300~s \\* 
BD $+$62 959     & LVMslits & \citet{Fulbright00} & \phn\phn1 & 2016 & Dec & 30 & 1.35 & $> 1\farcs5$ & 300~s \\*
BD $+$80 245     & LVMslits & \citet{Fulbright00} & \phn\phn1 & 2016 & Dec & 30 & 1.99 & $> 1\farcs5$ & 2 $\times$ 600~s \\* 
BD $+$21 1969     & LVMslits & \citet{Fulbright00} & \phn\phn1 & 2016 & Dec & 30 & 1.12 & $> 1\farcs5$ & 300~s \\*
BD $-$20 2955    & LVMslits & \citet{Fulbright00} & \phn\phn1 & 2016 & Dec & 30 & 1.35 & $> 1\farcs5$ & 300~s \\* 
BD $+$55 1362     & LVMslits & \citet{Fulbright00} & \phn\phn1 & 2016 & Dec & 30 & 1.26 & $> 1\farcs5$ & 2 $\times$ 300~s \\* 
BD $+$54 1359     & LVMslits & \citet{Fulbright00} & \phn\phn1 & 2016 & Dec & 30 & 1.28 & $> 1\farcs5$ & 2 $\times$ 300~s \\*
BD $+$40 2408     & LVMslits & \citet{Fulbright00} & \phn\phn1 & 2016 & Dec & 30 & 1.08 & $> 1\farcs5$ & 300~s \\*
BD $-$04 3155     & LVMslits & \citet{Fulbright00} & \phn\phn1 & 2016 & Dec & 30 & 1.32 & $> 1\farcs5$ & 300~s \\*
BD $+$49 2098     & LVMslits & \citet{Fulbright00} & \phn\phn1 & 2016 & Dec & 30 & 1.18 & $> 1\farcs5$ & 300~s \\*
BD $+$09 2653     & LVMslits & \citet{Fulbright00} & \phn\phn1 & 2016 & Dec & 30 & 1.25 & $> 1\farcs5$ & 300~s \\*
\cutinhead{dSphs}
Sculptor\tablenotemark{b}         & bscl1 & \citet{Kirby09}     &    \phn86 & 2011 & Jul & 31  & 1.72 & $0\farcs75$    & 4 $\times$ 1200~s, 900~s \\*
                 & bscl2 & \citet{Kirby09}      &       106 & 2011 & Aug & 6  & 1.82 & $0\farcs74$    & 2 $\times$ 1200~s, 2 $\times$ 840~s \\*
                 & bscl6 & \citet{Kirby09}      &    \phn91 & 2011 & Aug & 4  & 1.72 & $0\farcs83$    & 4 $\times$ 1200~s \\
Fornax\tablenotemark{b}           & bfor6 & \citet{Kirby10}      &       169 & 2011 & Aug & 5 & 1.86 & $1\farcs26$    & 2 $\times$ 800~s \\*
                 &          &           & & 2011 & Aug & 6 & 1.83 & $0\farcs76$    & 1200~s \\
                 &          &           & & 2011 & Aug & 7 & 1.90 & $0\farcs83$    & 3 $\times$ 1200~s \\
Sextans          & bsex2 & \citet{Kirby10}     &    \phn85 & 2016 & Jan & 30 & 1.10 & \nodata    & 2 $\times$ 1800~s \\*
                 &          &           & & 2016 & Jan & 31 & 1.11 & \nodata    & 2 $\times$ 1800~s, 1000~s \\
                 & bsex3 & \citet{Kirby10}     &    \phn85 & 2016 & Jan & 29 & 1.55 & \nodata    & 6 $\times$ 1550~s \\*
Ursa Minor\tablenotemark{b}       & bumi1 & \citet{Kirby10}      &       125 & 2011 & Jul & 29 & 1.52 & $0\farcs57$    & 600~s, 4 $\times$ 1200~s \\*
                 & bumi2 & \citet{Kirby10}      &       134 & 2011 & Jul & 31 & 1.68 & $0\farcs73$    & 4 $\times$ 1200~s \\*
                 & bumi3 & \citet{Kirby10}      &       137 & 2011 & Aug & 4 & 1.80 & $0\farcs64$    & 4 $\times$ 1200~s \\*
Draco\tablenotemark{b}            & bdra1 & \citet{Kirby10}      &       151 & 2011 & Jul & 30 & 1.42 & $1\farcs18$    & 5 $\times$ 1200~s \\*
                 & bdra2 & \citet{Kirby10}      &       167 & 2011 & Aug & 7 & 1.28 & $0\farcs67$    & 4 $\times$ 1200~s \\*
                 & bdra3 & \citet{Kirby10}      &       140 & 2011 & Aug & 5 & 1.37 & $0\farcs98$    & 5 $\times$ 1200~s 
\enddata
\tablenotetext{a}{This is the reference for the slitmask design (when applicable), star selection and membership verification.}
\tablenotetext{b}{These observations were originally published in \citet{Kirby15}.}
\end{deluxetable*}
\addtocounter{table}{1}

\subsection{Star Selection and Member Verification}

We relied on the star selection, member verification, and stellar parameters found in the literature. The details of how these stars were selected and how membership was determined are found in the references listed in Table \ref{tab:obs}. Here we outline these methods in two groups, which are separated by the concentration of the desired stars: targets observed with A) multi-object or B) single-object slitmasks.

A) Multi-object slitmasks were used to observe all dwarf galaxies and globular clusters. In general, multi-object slitmasks were designed using the dsimulator software package\footnote{https://www2.keck.hawaii.edu/inst/deimos/dsim.html}. Stars were prioritized to be included on the slitmask based on their overall brightness and the likelihood they are on the red giant branch---determined using their surface gravity (\logg) and position on the color--magnitude diagram. The stars were then verified as members of the galaxy or globular cluster based on radial velocity measurements. 

B) Single slit spectroscopy was used to obtain observations of Milky Way halo stars. Halo stars were selected from \citet{Fulbright00}, who verified that the stars belonged to the Milky Way halo using proper motion measurements. To exclude dwarf stars from our sample, we applied an additional constraint by only including stars with \logg\,$< 3.6$, adopting \citeauthor{Fulbright00}'s measurements of surface gravity.

\subsection{Spectroscopic Configuration and Reduction}
\label{sec:reduction}

Previous work (e.g., \citealt{Kirby10}) utilized the 1200 lines mm$^{-1}$ DEIMOS grating from 6400--9000\,\AA\ (with a spectral resolution of $R \approx 6500$ at $8500$\,\AA\ with $0\farcs7$ slits). However, the optical barium absorption lines are bluer (4554--6497\,\AA) than can be observed efficiently with this grating, which has a blaze wavelength of 7760\,\AA.\ Therefore, to measure individual barium abundances for each star, we chose a lower resolution grating that allowed bluer wavelengths to be observed. We used the 900ZD DEIMOS grating (900\,lines mm$^{-1}$), which has a blaze wavelength of 5500\,\AA.\ This configuration can yield up to 80--150 spectra per slitmask with medium resolution ($\approx 1.96$\,\AA\ FWHM or R $\approx 2550$ at $5000$\,\AA\ with $0\farcs7$ slits). A central wavelength of $5500$\,\AA\ coupled with an order-blocking filter (GG400) results in a spectral range of 4000--7200\,\AA.\ Kr, Ne, Ar, and Xe arc lamps were used for wavelength calibration, and a quartz lamp was used for flat fielding.

We reduced all observations using the \texttt{spec2d} pipeline \citep{Newman13,Cooper12}. This pipeline automatically determines the wavelength solution using the spectral arcs. However, these wavelength solutions were not sufficiently accurate, because the arc lamps have few detectable emission lines with wavelengths less than $\approx5000$\,\AA\ at the exposure times that we used. To improve the wavelength solution, synthetic spectra were generated and cross-correlated against the observed spectra at H$\alpha, H\beta,$ and H$\gamma$ in windows of 20\AA.\ A line was fit to establish the wavelength correction ($\Delta\lambda$) as a function of wavelength ($\lambda$). This solution precludes the ability to measure absolute radial velocities, but our only focus for this work is to measure barium abundances. 

We corrected for the global continuum by first fitting a spline to the observed spectrum with a break point every 200 pixels (88\,\AA) with an upper and lower threshold of $5\sigma$ and $0.1\sigma$, respectively. Then, we divided the observed spectrum by the resulting spline fit to correct for the global continuum. We later refine the local continuum determination during the barium abundance measurements (Section~\ref{sec:ba_measurement}).

\section{Barium Abundance Measurements}

We measure barium abundances by first adopting stellar parameter measurements found in the literature, specifically the effective temperature (\teff), surface gravity (\logg), metallicity (\feh), and $\alpha$-to-iron ratio (\afe). Abundances presented in this paper are referenced to solar (e.g., \feh\,$= ($Fe/H$)_\ast - ($Fe/H$)_{\odot}$). Our definition of the solar elemental abundances can be found in Table \ref{sun}. 

\setcounter{table}{2}
\begin{deluxetable}{lc}
\tablecolumns{2}
\tablecaption{Adopted Solar Composition\label{sun}}
\tablehead{\colhead{Element} & \colhead{(X/H) = 12 + $\log \epsilon_X$}} 
\startdata
Mg & 7.58\\
Fe & 7.52\\
Ba & 2.13\\
Eu & 0.51
\enddata
\tablecomments{Solar abundances are from \citet{Anders89}, except for iron which is from \citet{Sneden92}. Elemental abundance is defined as: (X/H)\,$=12 + \log\epsilon_X = 12 + \log(n_X) - \log(n_H)$.}
\end{deluxetable}

For dwarf galaxies and globular clusters, we adopted the parameter measurements from \citet{Kirby10} and \citet{Kirby16}, respectively. See \citet{Kirby10,Kirby16} for a full description of those measurements, which we briefly summarize here. The surface gravity and initial value of \teff\ were estimated from photometry. Then \teff, \feh, and \afe\ were measured by matching the synthetic spectra to the observed spectra. The microturbulent velocity ($\xi$) of the stellar atmosphere was calculated from the surface gravity (Equation \ref{microturbulence})\footnote{This formula for the microturbulent velocities is described by \citet{Kirby09}, who derived it by fitting spectroscopically measured microturbulent velocities and surface gravities from red giant branch stars in globular clusters in the literature.}. 

\begin{equation}
\label{microturbulence}
\xi(\text{km\,s}^{-1})=(2.13\pm0.05)-(0.23\pm0.03)\log g
\end{equation}

For Milky Way halo stars, we adopted \teff, \logg, \feh, and \afe\ found by \citet{Fulbright00}. As with the globular cluster and dwarf galaxy stars, the \afe\ used is the average abundance measured for $\alpha$ elements (i.e., Mg, Si, Ca, and Ti). We replaced the microturbulent velocity published by \citet{Fulbright00} with the results from Equation \ref{microturbulence}. We shifted the spectra to the rest frame based on the radial velocity measurements found by \citet{Fulbright02} before fine-tuning the wavelength solution using the Balmer lines (as described in Section \ref{sec:reduction}).

Now that we have discussed the sources of \teff, \logg, \feh, and \afe, we will describe how we measure \bafe\ from our DEIMOS observations using synthetic spectra.

\subsection{Synthetic Spectra}
\label{sec:synth}

Synthetic spectra were calculated for each combination of stellar parameters to measure \bafe\ from $-2.0$ to 1.0\,dex. Table \ref{grid} outlines our spectral grid by listing the range and step size for each parameter used to generate the synthetic spectra. For stars that had \bafe\ above 1.0\,dex or below $-2.0$\,dex, we computed additional synthetic spectra as needed. 

\begin{deluxetable}{lccr}
\tablewidth{0pt}
\tablecolumns{4}
\tablecaption{Barium Synthetic Spectra Grid\label{grid}}
\tablehead{\colhead{Parameter} & \colhead{Minimum} & \colhead{Maximum} & \colhead{Step}\\\colhead{} & \colhead{Value} & \colhead{Value} & \colhead{}} \startdata
\teff\,(K) & 3500 & 5600 & 100\\
 & 5600 & 8000 & 200\\
\logg\,($g$ in cm\,s$^{-2}$) & 0.0 (\teff$ < 7000$\,K) & 5.0 & 0.5\\
 & 0.5 (\teff$\geq 7000$\,K) & 5.0 & 0.5\\
\feh & $-5.0$ & 0.0 & 0.1 \\
\afe & $-0.8$ & 1.2 & 0.1\\
\bafe & $-2.0$ & 1.0 & 0.1
\enddata
\end{deluxetable}

To measure \bafe, we only needed short segments (20\,\AA) of synthetic spectra centered at five optical barium absorption lines: 4554.0, 4934.2, 5853.7, 6141.7, and 6496.9\,\AA.\ 

MOOG \citep[a spectral synthesis code,][]{Sneden73,Sneden12} generates synthetic spectra for a set of parameters assuming local thermal equilibrium (LTE). We modified the 2014 version of MOOG to reduce the computation time by stripping out all functionality except the spectral synthesis routine and by parallelizing it. MOOG relies on a stellar atmosphere model and a list of atomic and molecular absorption lines (i.e., line list).

We used ATLAS9 \citep{Kurucz93}, a collection of 1-dimensional plane-parallel stellar atmosphere models, which were interpolated to match our fine spectral grid (see \citealt{Kirby09} for more details). Stellar atmosphere models with matching \afe\ were used to calculate the synthetic spectra because $\alpha$ elements are a significant source of free electrons, which affect the opacity and therefore the atmospheric structure.

\subsubsection{Line List}
The line list used to calculate the synthetic spectra was compiled from a few different sources. The bulk of the line list was generated using the Vienna Atomic Line Database \citep[VALD,][]{Ryabchikova15}, which includes atomic lines and CH, MgH, SiH, and C$_2$ molecular lines. In addition, CN lines were included from \citet{Sneden14}. To avoid unnecessary computation time, we only included lines of neutral and singly ionized species with excitation potentials less than 10\,eV and oscillator strengths (\loggf) greater than $-5.0$. The format of the line list can be seen in Table \ref{tab:linelist}. For each absorption line, the species, wavelength, excitation potential, and \loggf\ are listed.

\begin{deluxetable}{lccr}
\tablecolumns{4}
\tablecaption{Spectral Line List\label{tab:linelist}}
\tablehead{\colhead{Species} & \colhead{Wavelength (\AA)} & \colhead{Excitation Potential (eV)} & \colhead{\loggf}} 
\startdata
V$\,${\sc i}   & 4543.0096  &    2.7080  &  -2.712\\
Sc$\,${\sc i}  & 4543.0282  &    2.2957  &  -4.032\\
Cr$\,${\sc i}  & 4543.0796  &    5.2394  &  -2.972\\
CN  & 4543.0821  &    0.4159  &  -4.012\\
C$_2$ & 4543.0891  &    1.9158  &  -4.781\\
CN  & 4543.1178  &    0.9947  &  -2.429\\
CN  & 4543.1190  &    0.9947  &  -2.583\\
CN  & 4543.1226  &    0.9947  &  -3.730\\
Ti$\,${\sc i}  & 4543.1393  &    3.4238  &  -3.781\\
S$\,${\sc i}   & 4543.1789  &    9.4169  &  -2.534
\enddata
\tablecomments{The first ten lines of the line list are shown here. The line list is published in its entirety in the correct format for MOOG \citep{Sneden73,Sneden12}
  in the machine-readable format.}
\end{deluxetable}

The line list was calibrated against spectra of the Sun and Arcturus to ensure the synthetic spectra match high quality (R$\approx$150,000 with S/N$\approx$1,000) observed spectra. The observed Arcturus and solar spectra are both from \citet{Hinkle00}\footnote{ftp.noao.edu/catalogs/arcturusatlas/visual/}. The synthetic spectra were calculated (using MOOG and ATLAS9 stellar atmosphere models) with the following stellar parameters: Sun ($\text{ \teff}=5777\text{\,K, \logg}=4.44\text{, \feh}=0\text{, \afe}=0$), and Arcturus ($\text{\teff}=4286\text{\,K, \logg}=1.66\text{, \feh}=-0.52\text{, \afe}=0.26$, \citealt{Ramirez11}). The oscillator strengths of some of the lines in the line list were adjusted to better match the synthetic spectra to the observations. After the calibration was finished, the standard deviation of the absolute difference of the observed and synthetic flux is less than 4\%. We smoothed the spectra to match the resolution of the DEIMOS observations and found that the dispersion decreased to less than 1\%. The accuracy of the line list other than in the immediate region of the barium lines is only important for correcting the local continuum. For this purpose, 1\% agreement assures that the line list will not be the dominant source of error.

We adopted \citeauthor{McWilliam98}'s (\citeyear{McWilliam98}) line list for the five strong barium lines. This list accounts for hyper-fine and isotope splitting. We adopted the solar system barium isotope ratios from \citet{Anders89}. The impact of this assumption was tested by measuring the change in \bafe\ when assuming pure \rproc\ and \sproc\ isotope ratios from \citet{Sneden08} in eight stars from Sculptor that spanned the stellar parameters probed. We found a maximum change in \bafe\ of 0.04\,dex and an average change of 0.02\,dex and 0.008\,dex for \rproc\ and \sproc\ ratios, respectively. Compared to the measurement uncertainties in our barium measurements  ($\approx0.2$\,dex), isotope ratios are not a significant source of error.
 
\subsection{Barium Measurement Technique}
\label{sec:ba_measurement}

\begin{figure*}
\centering
\includegraphics[width=0.9\linewidth]{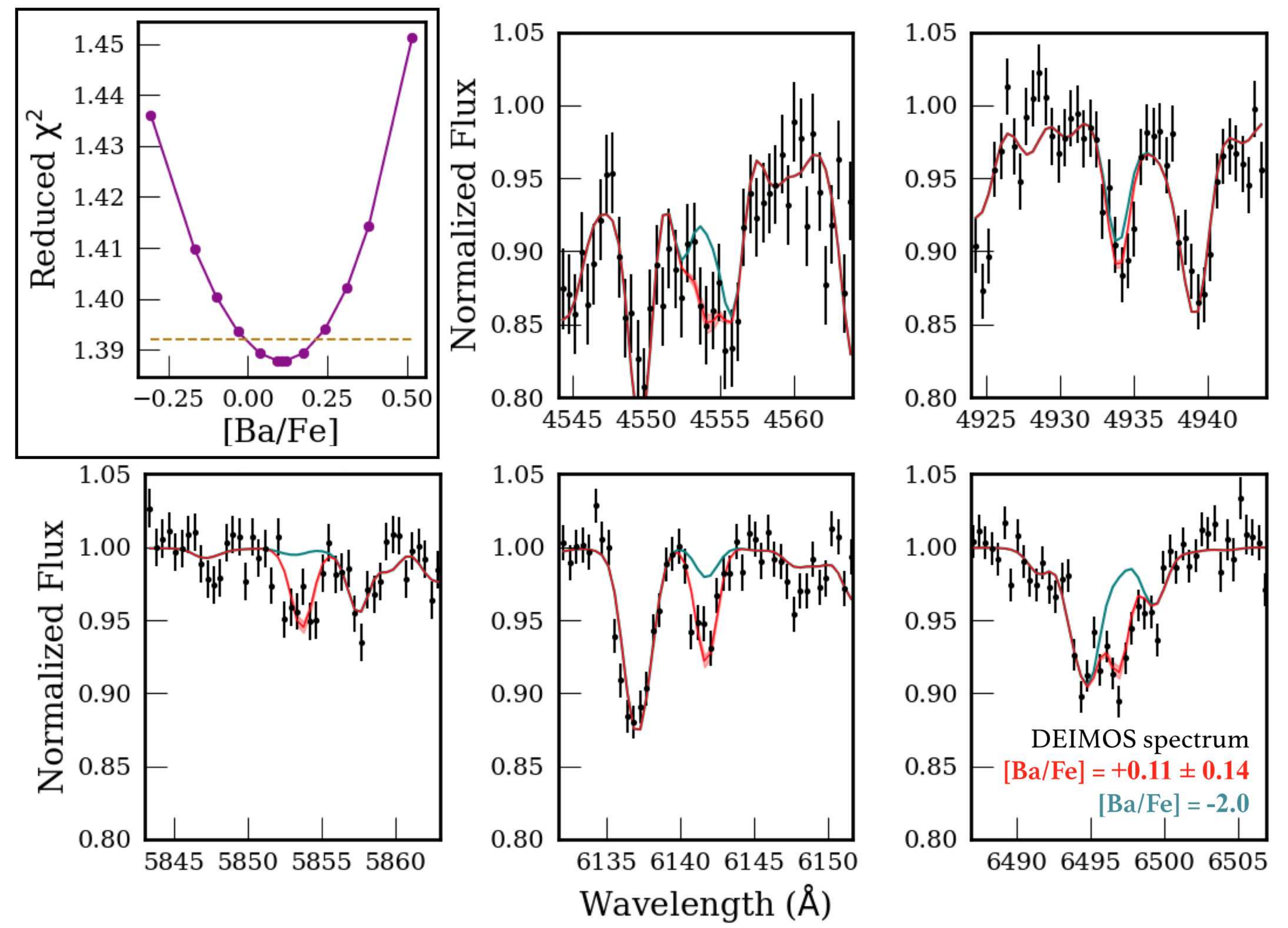}
\caption{Example barium measurement for a single star in Draco. A DEIMOS medium-resolution spectrum (black) is shown alongside two synthetic spectra: the best-fit synthetic spectrum with the statistical error ranged shaded ($\text{\bafe} = +0.11\pm0.14$\,dex, red) and one with $\text{\bafe}=-2.0$\,dex (blue). The top left panel shows the quality of the fit by comparing the reduced $\chi^2$ for synthetic spectra at a range of \bafe\ values (purple). The horizontal dotted brown line indicates the one sigma uncertainty of the measurement. While most individual absorption lines are not highly significant, the simultaneous fit of five barium lines enables uncertainties of $\approx 0.2$\,dex. 
\label{ba_fit}}
\end{figure*}

We interpolated the synthetic spectra from our grid to match the exact parameters published for a given star (\teff, \logg, \feh, and \afe) and smoothed them to match the resolution of the observed spectrum ($\sigma=0.73$\,\AA\ Gaussian kernel). The local continuum was corrected by fitting a line to each 20\,\AA\ segment centered on a barium line with 1\,\AA\ on either side of the barium line masked out. The optimal barium abundance was measured by matching the synthetic spectra for various values of \bafe\ to the observed spectrum. The best match was determined using a Levenberg-Marquardt fitting algorithm (via scipy.optimize.curve\textunderscore{}fit, \citealt{Jones01}). All barium lines are fit simultaneously, which is a key reason why we are able to measure accurate barium abundances in spectra with relatively low spectral resolution and S/N. If a given line doesn't provide a very useful constraint, due to noise or other issues, the fit relies on the other clearer lines. Most stars are measured using five barium lines, but occasionally the 5853.7\,\AA\ barium line falls in the DEIMOS chip gap, resulting in four barium lines being used. Figure \ref{ba_fit} demonstrates a \bafe\ fit for a single star. The top left panel shows the reduced chi-squared as a function of \bafe. The remaining five panels each display a wavelength segment centered on a barium absorption line. The \bafe\ error quoted in this figure does not include the systematic error. See Section \ref{sec:error} for a discussion on how the systematic uncertainty in \bafe\ is determined. The figure demonstrates that the barium measurement is well constrained and has a statistical error similar to high-resolution studies.

\subsection{The Catalog}
After we measured all the member stars in our sample, our catalog contains barium abundance measurements of 243 stars belonging to five dwarf galaxies, making this the largest self-consistent sample of dwarf galaxy barium abundances measured to date. Table \ref{tab:catalog} gives the multi-element abundance catalog for dwarf spheroidal galaxy (dSph) stars where \bafe\ has been measured with uncertainties less than 0.28\,dex. One advantage of this catalog of barium abundances is that this large sample of stars are all measured using the same assumptions in an automated way. This means that the catalog is internally consistent. The next section shows the consistency we achieve when comparing our \bafe\ measurements to other measurements found in the literature, but it should be kept in mind that those literature measurements entail heterogeneous observations and measurement techniques. 

\renewcommand{\thetable}{\arabic{table}}
\begin{deluxetable*}{llcccccccc}
\tablecolumns{10}
\tablecaption{DEIMOS Multi-Element Abundances Catalog of dSph Stars\label{tab:catalog}}
\tablehead{\colhead{dSph} & \colhead{Name} & \colhead{RA} & \colhead{Dec} & \colhead{\teff} & \colhead{\logg} & \colhead{$\xi$} & \colhead{\feh} & \colhead{\afe} & \colhead{\bafe} \\
\colhead{ } & \colhead{ } & \colhead{ } & \colhead{ } & \colhead{(K)} & \colhead{($g$ in cm\,s$^{-2}$)} & \colhead{(km\,s$^{-1}$)} & \colhead{(dex)} & \colhead{(dex)} & \colhead{(dex)}}
\startdata
Dra & 615574 & 17h20m07.44s & +57d54m32.7s & $4767\pm46$ & 1.44 & 1.80 & $-1.90\pm0.10$ & $-0.30\pm0.18$ & $+0.01\pm0.24$ \\
Dra & 622253 & 17h19m57.29s & +57d55m04.6s & $4536\pm23$ & 1.14 & 1.87 & $-1.36\pm0.10$ & $-0.18\pm0.09$ & $+0.63\pm0.13$ \\
Dra & 649595 & 17h20m01.59s & +57d57m04.6s & $4500\pm34$ & 0.87 & 1.94 & $-1.89\pm0.10$ & $-0.04\pm0.10$ & $+0.04\pm0.16$ \\
Dra & 648189 & 17h19m57.91s & +57d56m58.4s & $4479\pm27$ & 0.81 & 1.95 & $-2.16\pm0.10$ & $+0.11\pm0.09$ & $-0.16\pm0.16$ \\
Dra & 588559 & 17h20m13.51s & +57d51m59.3s & $4447\pm23$ & 0.77 & 1.96 & $-1.97\pm0.10$ & $+0.00\pm0.09$ & $-0.56\pm0.15$ \\
Dra & 660453 & 17h20m05.66s & +57d57m52.8s & $4363\pm24$ & 0.73 & 1.97 & $-2.01\pm0.10$ & $+0.05\pm0.09$ & $+0.16\pm0.15$ \\
Dra & 598482 & 17h20m16.12s & +57d52m56.1s & $4418\pm49$ & 0.77 & 1.96 & $-3.02\pm0.10$ & $+0.14\pm0.18$ & $-1.39\pm0.24$ \\
Dra & 676918 & 17h20m03.97s & +57d59m08.3s & $4255\pm16$ & 0.40 & 2.05 & $-1.66\pm0.10$ & $-0.08\pm0.09$ & $+0.09\pm0.13$ \\
Dra & 640120 & 17h20m05.78s & +57d56m23.4s & $4419\pm27$ & 0.50 & 2.02 & $-2.21\pm0.10$ & $+0.19\pm0.09$ & $-0.09\pm0.13$ \\
Dra & 653393 & 17h19m58.88s & +57d57m20.9s & $4242\pm20$ & 0.12 & 2.11 & $-2.41\pm0.10$ & $+0.00\pm0.09$ & $-1.30\pm0.14$ 
\enddata
\tablecomments{Table~\ref{tab:catalog} is published in its entirety in the machine-readable format. A portion is shown here for guidance regarding its form and content. The errors reported here already include the systematic errors for \feh, \afe, and \bafe\ found in Table~\ref{tab:syserr}. Stars are only included if the errors for \feh, \afe, and \bafe\ are less than 0.28\,dex.}
\end{deluxetable*}
      
\section{Systematic Uncertainty}
\label{sec:error}

Beyond the \bafe\ statistical errors returned by our fitting algorithm, we need to consider the systematic error of our measurement. The \bafe\ statistical errors reflect the noise in the spectrum and how precisely the synthetic spectrum matches the observed spectrum. Technically, the statistical error reported is the square root of the diagonal values of the covariance matrix generated by Python's scipy.optimize.curve\_fit \citep{Jones01} function when fitting synthetic spectra to the observed spectrum. The \bafe\ systematic error could be caused by the assumptions used in the spectral synthesis code (e.g., non-LTE effects), details of the method used to measure abundances, the line list, and how errors in other stellar parameters impact the \bafe\ measured. The contribution of these sources help us to establish an error floor that is added in quadrature with the statistical errors to produce the errors reported in the catalog.

\subsection{Appropriateness of the LTE Assumption}
\label{sec:nonLTE}

We use a spectral synthesis code that assumes local thermal equilibrium (LTE), because it greatly simplifies the computational burden. In the LTE case the opacity needs be known as a function of only temperature and density to solve for the flux.
However, the impact of assuming LTE is perhaps the most significant assumption of all the assumptions made in the spectral synthesis code. Assuming LTE is valid only when the radiation field is closely coupled to the matter, which occurs through collisions
between atoms and electrons. Therefore LTE holds at high densities.

A couple studies have carefully measured barium without assuming LTE (i.e., non-LTE) and compared them to LTE abundances published in the literature. For example, \citet{Andrievsky09, Andrievsky17} found that the difference between \bafe\ measured with and without assuming LTE varies from negligible to very significant ($\lesssim 0.8$\,dex). The impact of assuming LTE on barium abundance measurements is primarily dependent on the \teff\ and \bah\ of the star in question \citep{Andrievsky09}. In addition, the details of the measurement (e.g., how many and which absorption lines are used) also plays a role on how sensitive the abundance measurements are to LTE effects. 

\begin{figure}
\centering
\includegraphics[width=\linewidth]{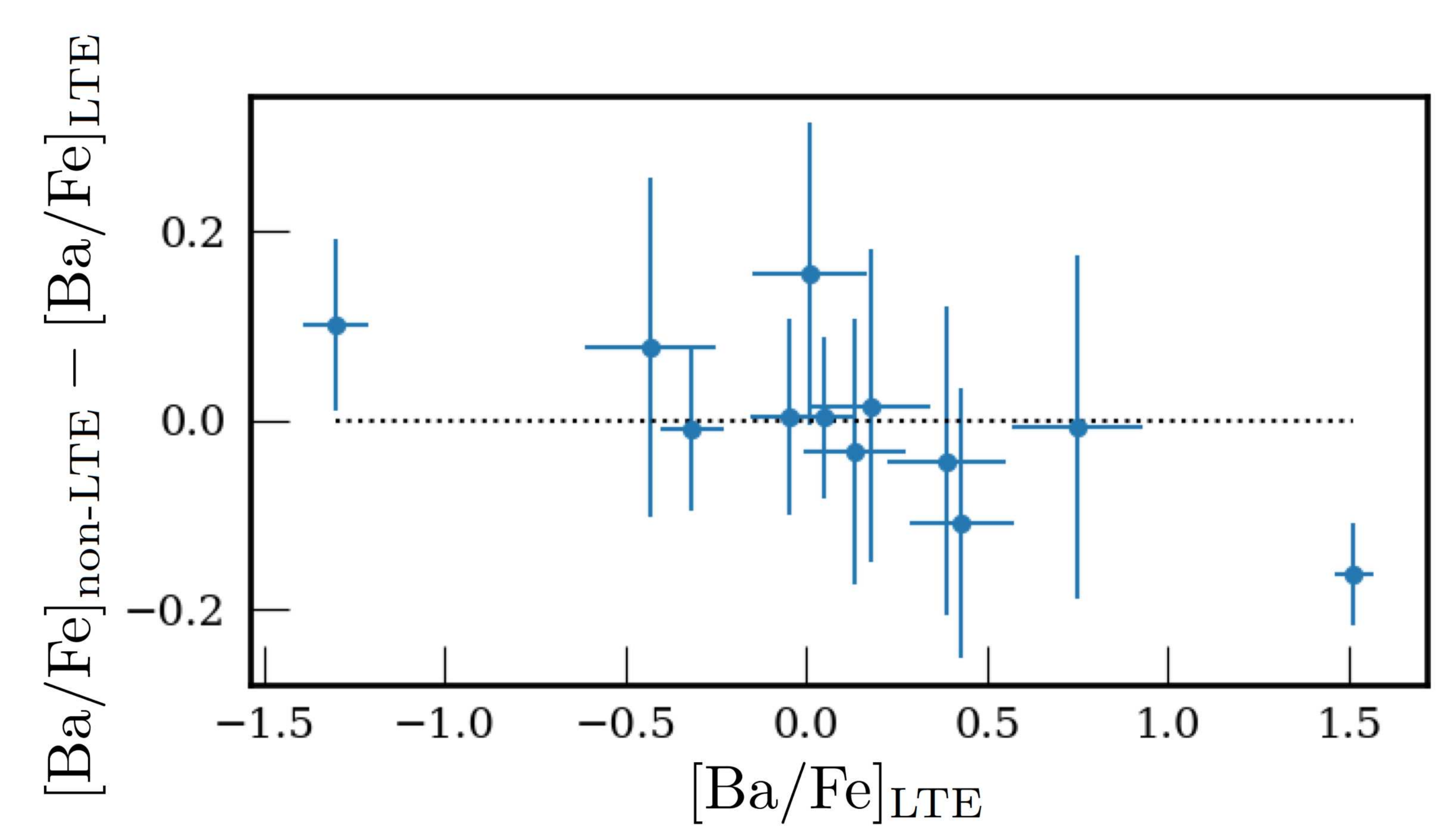}
\caption{Comparison between our \bafe\ measurements assuming LTE and without this assumption (i.e., non-LTE) using the same spectra and stellar parameters. These twelve representative stars show that the effect of assuming LTE is mostly within our statistical uncertainties.
\label{Ba_nlte_comp}}
\end{figure}

To test the impact of assuming LTE with our observations, we measured a small, representative subset of stars with and without assuming LTE. We selected twelve stars that spanned the range of \teff\ and \bah\ seen in our full sample. Both \bafe\ measurements used the same DEIMOS spectra and stellar parameters. \citet{Andrievsky17} describes the detailed methods of the non-LTE barium measurements. Our \bafe\ measurements assuming LTE are consistent with the non-LTE measurements, as seen in Figure \ref{Ba_nlte_comp}. The results are also reported in Table \ref{tab:non-lte}. Given that the impact of assuming LTE on \bafe\ is well within the statistical uncertainties, no additional systematic uncertainty is needed to account for the effect of assuming LTE.

\begin{deluxetable*}{llccccccc}
\tablecolumns{9}
\tablecaption{Non-LTE Effects\label{tab:non-lte}}
\tablehead{
\colhead{dSph} & \colhead{Name} &  \colhead{\teff} & \colhead{\logg} & \colhead{$\xi$} & \colhead{\feh} & \colhead{\afe} & \colhead{\bafe$_{\text{LTE}}$} & \colhead{\bafe$_{\text{non-LTE}}$} \\
\colhead{ } &  \colhead{ } & \colhead{(K)} & \colhead{($g$ in cm\,s$^{-2}$)} & \colhead{(km\,s$^{-1}$)} & \colhead{(dex)} & \colhead{(dex)} & \colhead{(dex)} & \colhead{(dex)} }
\startdata
For & 74926 & $3747\pm11$ & 0.21 & 2.09 & $-1.03\pm0.10$ & $-0.20\pm0.10$ & $+0.39\pm0.19$ & $+0.34$ \\
For & 64059 & $3794\pm12$ & 0.32 & 2.07 & $-1.05\pm0.10$ & $-0.21\pm0.10$ & $+0.01\pm0.19$ & $+0.16$ \\
For & 58973 & $3856\pm11$ & 0.36 & 2.06 & $-1.14\pm0.10$ & $-0.09\pm0.10$ & $+0.13\pm0.17$ & $+0.10$ \\
For & 55262 & $4016\pm19$ & 0.45 & 2.04 & $-0.91\pm0.10$ & $+0.32\pm0.13$ & $+0.43\pm0.17$ & $+0.32$ \\
Scl & 1018551 & $4182\pm20$ & 0.53 & 2.02 & $-1.72\pm0.10$ & $+0.14\pm0.09$ & $-0.05\pm0.14$ & $-0.05$ \\
Scl & 1008522 & $4230\pm14$ & 0.55 & 2.01 & $-2.01\pm0.10$ & $+0.11\pm0.09$ & $+0.05\pm0.13$ & $+0.05$ \\
Dra & 653393 & $4242\pm20$ & 0.12 & 2.11 & $-2.41\pm0.10$ & $+0.00\pm0.09$ & $-1.30\pm0.14$ & $-1.20$ \\
Scl & 1003537 & $4280\pm18$ & 0.58 & 2.01 & $-2.26\pm0.10$ & $+0.21\pm0.09$ & $-0.32\pm0.13$ & $-0.33$ \\
Scl & 1013218 & $4622\pm44$ & 1.49 & 1.79 & $-1.55\pm0.10$ & $-0.06\pm0.11$ & $+1.51\pm0.11$ & $+1.35$ \\
Scl & 1016539 & $4614\pm51$ & 1.40 & 1.81 & $-1.76\pm0.11$ & $+0.17\pm0.13$ & $+0.18\pm0.19$ & $+0.19$ \\
Scl & 1010313 & $4770\pm52$ & 1.61 & 1.76 & $-2.78\pm0.13$ & $+0.47\pm0.17$ & $-0.43\pm0.21$ & $-0.35$ \\
UMi & Pal119 & $5075\pm48$ & 2.00 & 1.67 & $-1.83\pm0.11$ & $-0.36\pm0.23$ & $+0.75\pm0.21$ & $+0.74$
\enddata
\end{deluxetable*}

\subsection{Comparison to Other Catalogs}
\label{sec:HRS_MRS}

We must account for the systematic error introduced by any inaccuracies in the synthetic spectra and in the determination of stellar parameters. One way to account for this is to compare \bafe\ measured for stars that overlap between our catalog and other catalogs published in the literature. All abundances have been shifted to the same solar abundance scale, which is described in Table~\ref{sun}. Table \ref{tab:hrsmethod} highlights the different methods used by each literature source. Some of the differences between the measurements reflect the diversity of measurement techniques. In the following we refer to the literature sources as HRS because they all utilize high-resolution spectra. An advantage of our MRS sample is the uniformity of the acquisition and analysis of the spectra. 

\begin{deluxetable*}{lcccccc}
\tablecolumns{7}
\tablecaption{Previously Published HRS Abundance Methods\label{tab:hrsmethod}}
\tablehead{\colhead{Reference} & \colhead{System} & \colhead{\phantom{\tablenotemark{a}}Atmospheres\tablenotemark{a}} & \colhead{\phantom{\tablenotemark{b}}Code\tablenotemark{b}} & \colhead{\phantom{\tablenotemark{c}}\teff\tablenotemark{c}} & \colhead{\phantom{\tablenotemark{d}}\logg\tablenotemark{d}} & \colhead{\phantom{\tablenotemark{e}}$\xi$\tablenotemark{e}}} 
\startdata
\cutinhead{Globular Clusters}
\protect \citet{Cohen12} & NGC 2419 & ATLAS9 & MOOG & phot  & phot & spec \\
\protect \citet{Worley13} & M15 & ATLAS9 & MOOG & phot & phot & \phantom{\tablenotemark{f}}spec\tablenotemark{f} \\
\protect \citet{Lee05} & M68 & ATLAS9 & MOOG & spec & \phantom{\tablenotemark{g}}spec\tablenotemark{g} & spec \\
\protect \citet{Venn12} & M68 & MARCS & MOOG & spec & phot & spec \\
\cutinhead{Halo Field Stars}
\protect \citet{Fulbright00} && ATLAS9 & MOOG & spec & spec & spec \\
\cutinhead{dSph}
\protect \citet{Aoki07} & Ursa Minor & ATLAS9 & \citet{Tsuji78}, \citet{Aoki09_synth} & spec  & spec & spec \\
\protect \citet{Cohen09,Cohen10} & Draco, Ursa Minor & ATLAS9 & MOOG & spec  & spec & spec \\
\protect \citet{Geisler05} & Sculptor & MARCS & MOOG & phot & spec & spec \\
\protect \citet{Letarte18} & Fornax & MARCS & CALRAI & phot & phot & spec \\
\protect \citet{Sadakane04} & Ursa Minor & ATLAS9 & SPTOOL & spec & spec & spec\\
\protect \citet{Shetrone01,Shetrone03} & many & MARCS & MOOG & spec & spec & spec \\
\protect \citet{Tsujimoto17} & Draco & ATLAS9 & \citet{Tsuji78}, \citet{Aoki09_synth} & phot & phot & spec \\
\enddata
\tablenotetext{a}{ATLAS9: \citet{Kurucz93}, \citet{Castelli04}, \protect \url{http://kurucz.harvard.edu/grids.html}; MARCS: \protect \citet{Gustafsson75,Gustafsson03,Gustafsson08}.}
\tablenotetext{b}{MOOG: \protect \citet{Sneden73}; CALRAI: \protect \citet{Spite67}; SPTOOL: \citet{Takeda95}; \citet{Tsuji78}, \citet{Aoki09_synth}.}
\tablenotetext{c}{phot: model isochrones or empirical color-\teff\ relation; spec: Fe~I excitation equilibrium.}
\tablenotetext{d}{phot: model isochrones or \teff, with assuming a stellar mass and determining the luminosity from bolometric corrections; spec: Fe~I and Fe~II ionization balance.}
\tablenotetext{e}{spec: removing abundance trends as a function of equivalent width.}
\tablenotetext{f}{A \logg--$\xi$ relation derived from a subset of stars was applied to the full sample.}
\tablenotetext{g}{Photometric values were also published, but we have adopted the abundances measured with the spectroscopic values.}
\end{deluxetable*}

\begin{figure}
\centering
\includegraphics[width=\linewidth]{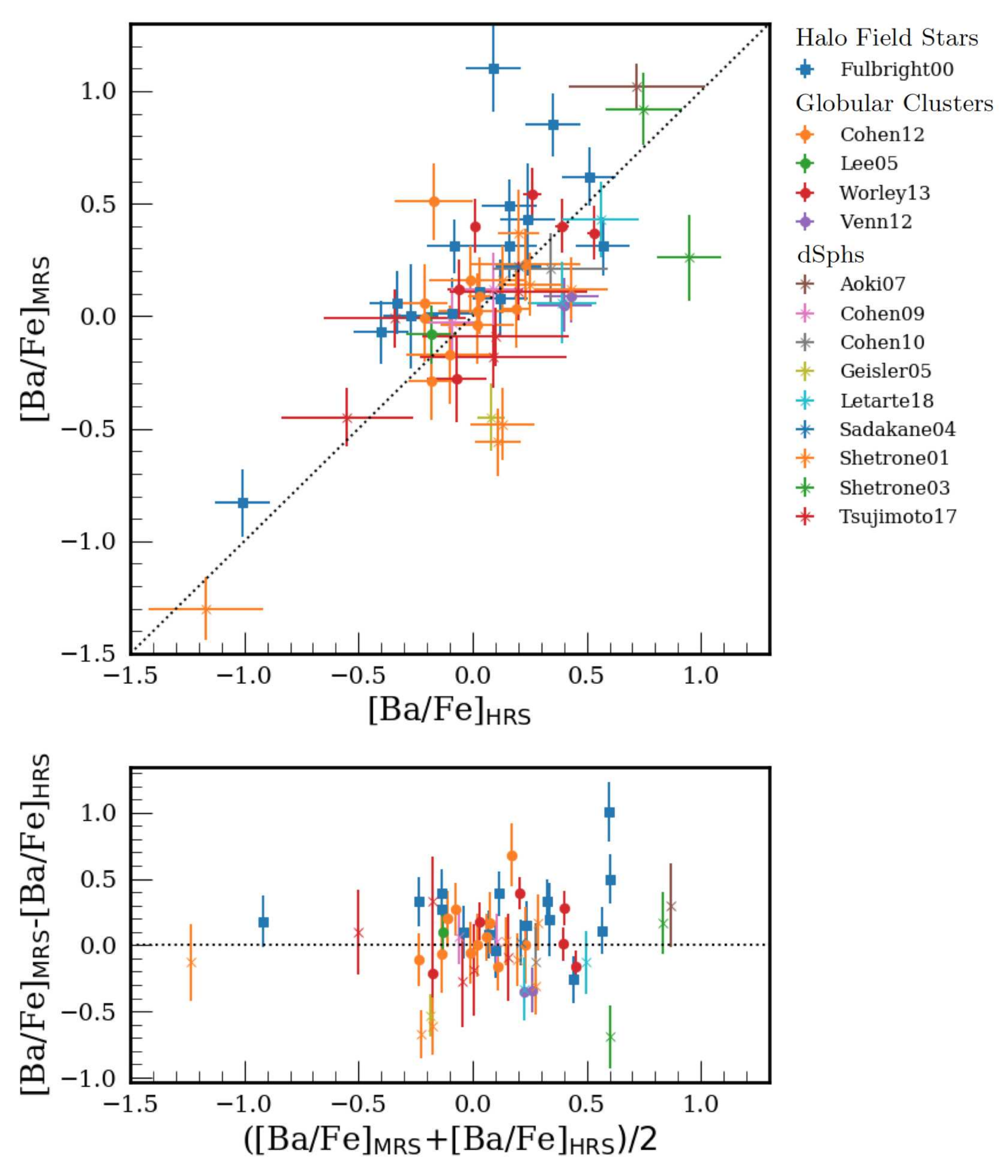}
\caption{Comparison of our medium-resolution spectroscopy (MRS) to high-resolution spectroscopy (HRS) abundance measurements for 57 stars. See the legend of this figure or Table~\ref{tab:hrscompare} for the sources of the HRS measurements.
\label{HRS_abund}}
\end{figure}

We compare a total of 57 stars that span \bafe\ values from $-1.2$ to 0.8\,dex. The stellar parameters and \bafe\ measurements for both MRS and HRS methods are contained in Table \ref{tab:hrscompare}. Figure \ref{HRS_abund} compares our barium abundances (\bafe$_{\text{MRS}}$) to the barium abundances published in the literature (\bafe$_{\text{HRS}}$). The barium abundances from MRS and HRS are largely consistent across the more than 1\,dex span of \bafe\ probed, but the difference ($\text{\bafe}_{\text{MRS}}-\text{\bafe}_{\text{HRS}}$) has a small average offset between the MRS and HRS \bafe\ measurements of 0.03\,dex. However, the difference ($\text{\bafe}_{\text{MRS}}-\text{\bafe}_{\text{HRS}}$) has a significant scatter.

\begin{deluxetable*}{lllcccc}
\tabletypesize{\scriptsize}
\tablecolumns{7}
\tablecaption{Comparison Between High-Resolution and DEIMOS Abundances\label{tab:hrscompare}}
\tablehead{\colhead{System} & \colhead{Name} & \colhead{HRS Reference} & \colhead{\feh$_{\text{MRS}}$} & \colhead{\bafe$_{\text{MRS}}$} & \colhead{\feh$_{\text{HRS}}$} & \colhead{\bafe$_{\text{HRS}}$} \\
\colhead{ } & \colhead{ } & \colhead{ } & \colhead{(dex)} & \colhead{(dex)}  &  \colhead{(dex)} & \colhead{(dex)}}
\startdata
NGC 2419 & N2419-S1004 & \citet{Cohen12} & $-2.15\pm0.09$ & $-0.17\pm0.22$ & $-2.15\pm0.14$ & $-0.10\pm0.19$ \\
NGC 2419 & N2419-S1065 & \citet{Cohen12} & $-2.11\pm0.09$ & $-0.01\pm0.19$ & $-2.10\pm0.15$ & $-0.21\pm0.09$ \\
NGC 2419 & N2419-S1131 & \citet{Cohen12} & $-2.12\pm0.08$ & $+0.09\pm0.17$ & $-2.10\pm0.15$ & $+0.03\pm0.05$ \\
NGC 2419 & N2419-S1166 & \citet{Cohen12} & $-2.04\pm0.08$ & $-0.29\pm0.17$ & $-2.07\pm0.13$ & $-0.18\pm0.10$ \\
NGC 2419 & N2419-S1209 & \citet{Cohen12} & $-2.29\pm0.08$ & $-0.04\pm0.17$ & $-2.32\pm0.11$ & $+0.02\pm0.16$ \\
NGC 2419 & N2419-S1305 & \citet{Cohen12} & $-2.21\pm0.09$ & $+0.03\pm0.17$ & $-2.25\pm0.18$ & $+0.19\pm0.06$ \\
NGC 2419 & N2419-S1814 & \citet{Cohen12} & $-2.20\pm0.08$ & $+0.16\pm0.15$ & $-2.26\pm0.14$ & $-0.01\pm0.18$ \\
NGC 2419 & N2419-S223 & \citet{Cohen12} & $-2.23\pm0.08$ & $+0.23\pm0.16$ & $-2.19\pm0.15$ & $+0.23\pm0.24$ \\
NGC 2419 & N2419-S406 & \citet{Cohen12} & $-2.10\pm0.09$ & $+0.06\pm0.17$ & $-2.10\pm0.13$ & $-0.21\pm0.10$ \\
NGC 2419 & N2419-S458 & \citet{Cohen12} & $-2.10\pm0.09$ & $+0.51\pm0.17$ & $-2.15\pm0.15$ & $-0.17\pm0.17$ \\
\enddata
\tablecomments{Table~\ref{tab:hrscompare} is published in its entirety in the machine-readable format.
      Some columns (e.g., \teff, \logg, $\xi$ and \afe\ for both MRS and HRS) are
  suppressed in the printed edition, and only the first ten lines are shown. Although the Milky Way halo field stars \citep{Fulbright00} were measured with the same stellar parameters (e.g., \teff, \logg, \feh, \afe) in both methods, stars in the dwarf galaxies and globular clusters have different stellar parameters in each method. Stars are only included if the errors for \feh, \afe, and \bafe\ are less than 0.28\,dex.}
\end{deluxetable*}

The standard deviation of the offset between the MRS and HRS \bafe\ measurements provides a metric to determine the systematic uncertainty. Assuming the uncertainties published in the HRS \bafe\ measurements are prefect representations of the true error, we solved for the systematic uncertainty required to standardize the offset. This is done by solving Equation \ref{eq:systematic_error} for the systematic error ($\sigma_{\text{sys}}$).

\begin{equation}
\label{eq:systematic_error}
\text{stddev}\left(\frac{\text{\bafe}_{\text{MRS}}-\text{\bafe}_{\text{HRS}}}{\sqrt{\sigma_{\text{HRS}}^2+\sigma_{\text{MRS}}^2+\sigma_{\text{sys}}^2}}\right) = 1
\end{equation}

\begin{figure}
\centering
\includegraphics[width=0.95\linewidth]{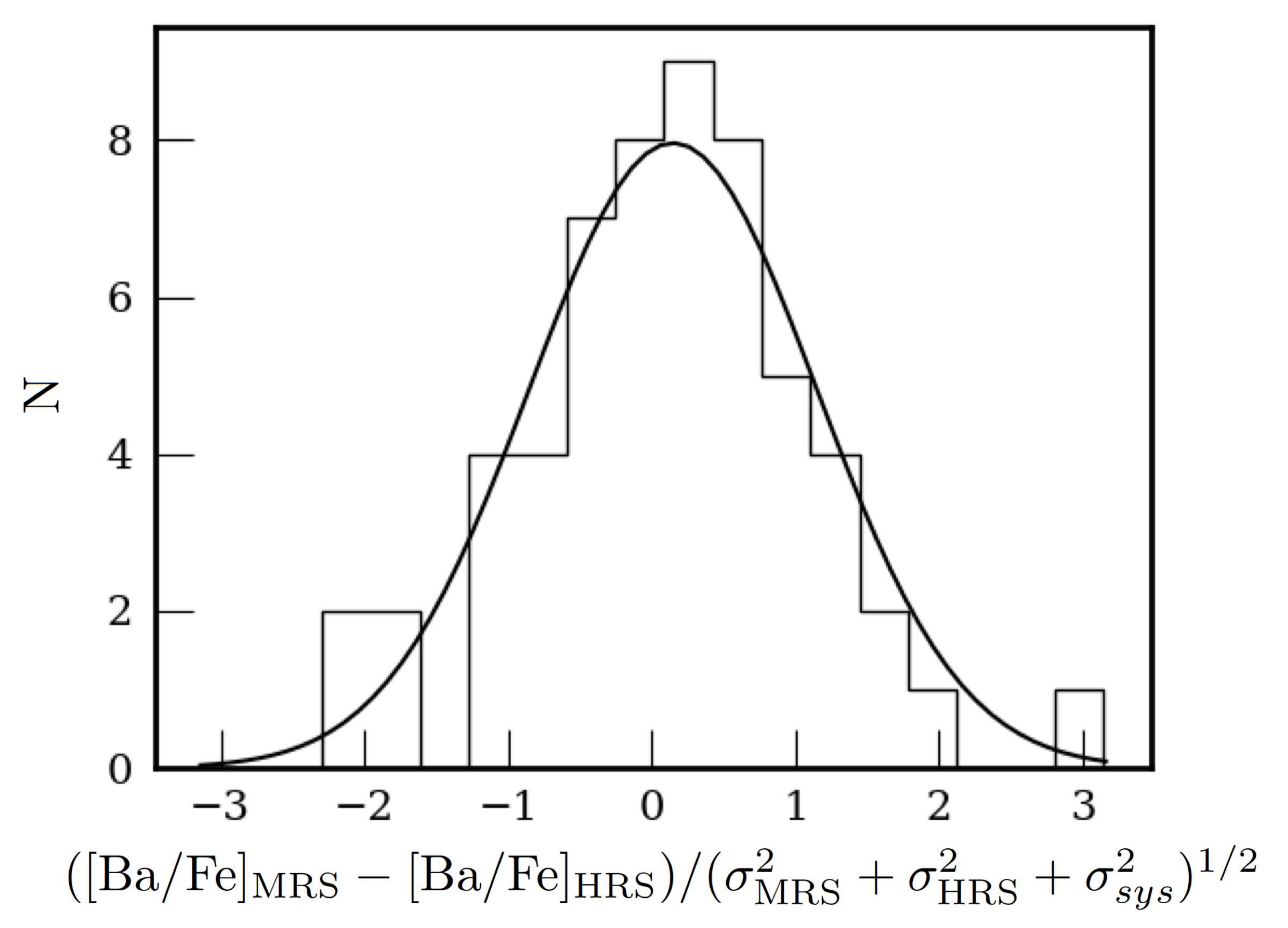}
\caption{Distribution of differences between our medium-resolution spectroscopy (MRS) and high-resolution spectroscopy (HRS) measurements of \bafe\ divided by the estimated error of the difference. The histogram contains 57 stars with a measured $\sigma_{\text{sys}}=0.23$. The curve is a Gaussian with $\sigma=1$.
\label{sys_hist}}
\end{figure}

\noindent Thus, $\sigma_{\text{sys}}$ is the error required to be added in quadrature with the MRS statistical error to force the dispersion between MRS and HRS to be unity. By comparing the MRS and HRS measurements of 57 stars, we measure $\sigma_{\text{sys}}=0.23$\,dex.
The resulting error distribution is shown in Figure~\ref{sys_hist}.  

\subsection{Intrinsic Dispersion}

\begin{figure}
\centering
\includegraphics[width=\linewidth]{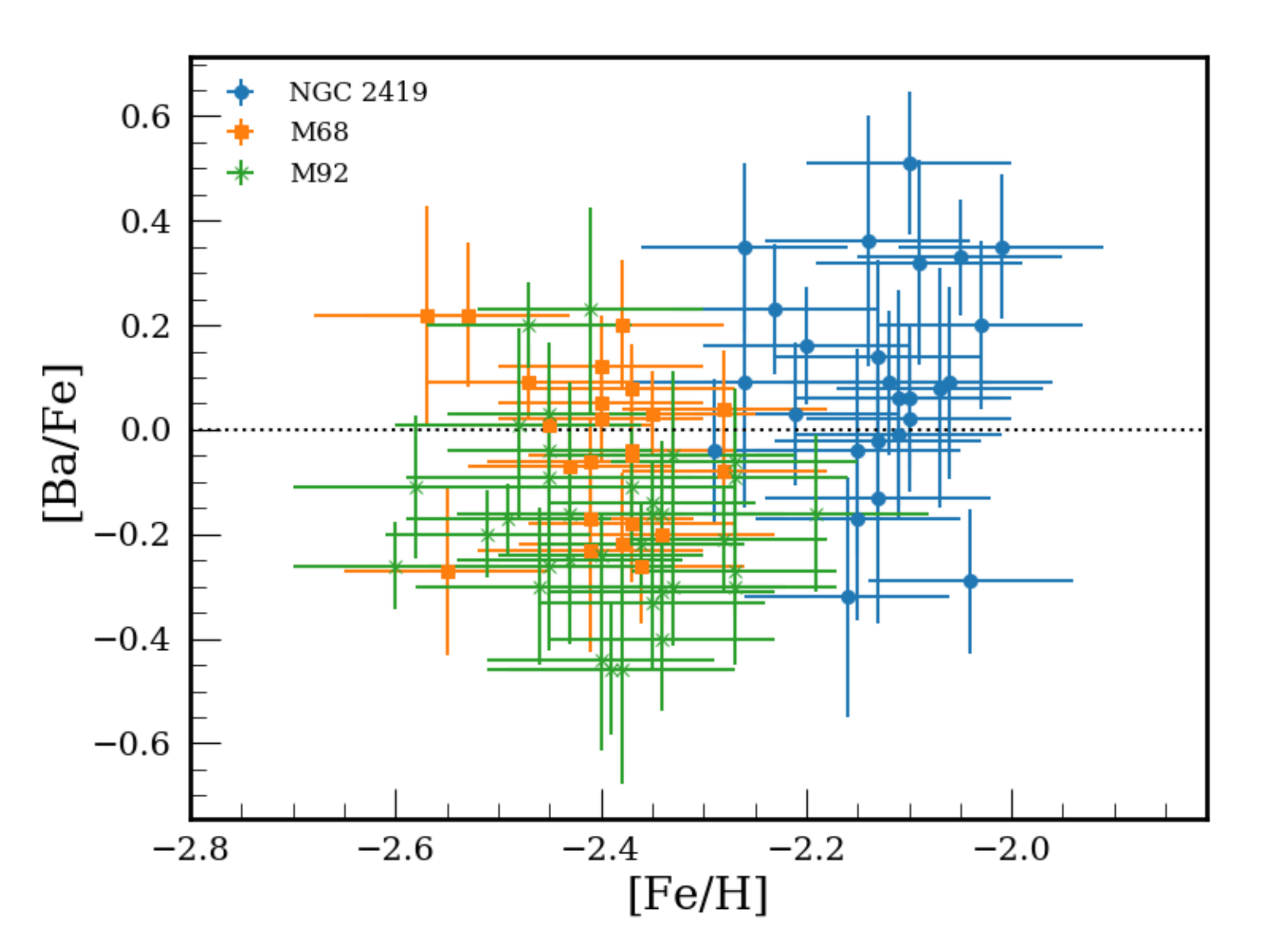}
\caption{We have measured \bafe\ for 26, 23, and 34 stars in NGC 2419, M68, and M92, respectively. This sample is used to constrain the systematic uncertainty of our measurement.
\label{M92}}
\end{figure}

\begin{figure}
\centering
\includegraphics[width=0.95\linewidth]{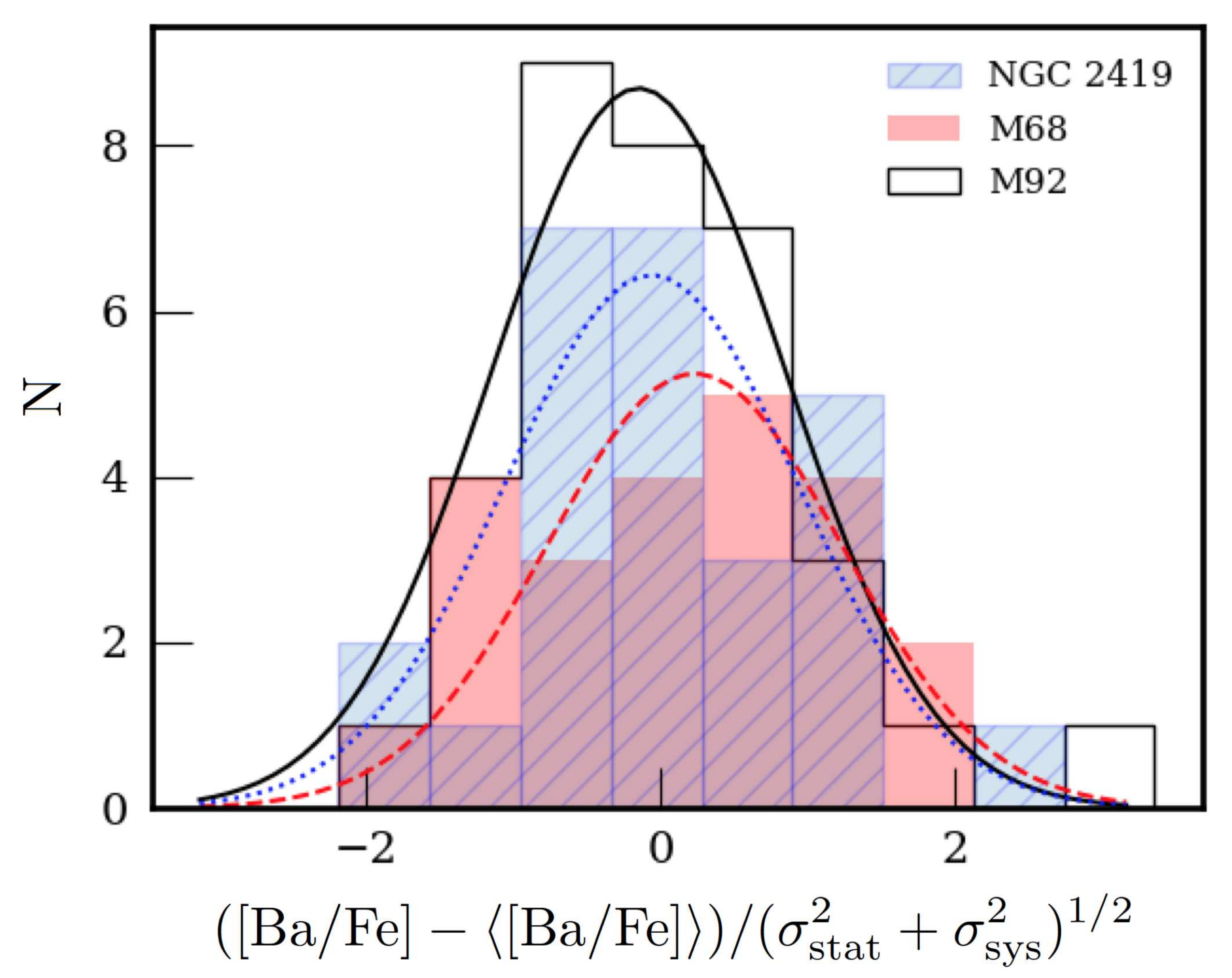}
\caption{Distribution of differences from the average \bafe\ measured in each globular cluster divided by the estimated error of the difference. By setting the intrinsic dispersion of NGC 2419, M68, and M92 to zero, we measured $\sigma_{\text{sys}}=0.11$, 0.07, and 0.09, respectively. The curves are Gaussians with $\sigma=1$.
\label{M92_sys_hist}}
\end{figure}

With some exceptions \citep[e.g., M15;][]{Sneden97}, most globular clusters are expected to have no internal variation in heavy elements. Here we measure the systematic error that would be required to ensure there is no intrinsic dispersion of \bafe\ in a globular cluster. This measurement was done with NGC 2419, M68, and M92 using Equation~\ref{eq:intrinsic_error}, where $\langle\text{\bafe}\rangle$ is the average abundance. This assumes that these globular clusters do not have intrinsic dispersion in \bafe, which may or may not be the case \citep[e.g.,][]{RoedererSneden11,Cohen11}, so the systematic uncertainty returned is an upper limit. By standardizing the offset (see Equation \ref{eq:intrinsic_error}) of the $\approx30$ stars for which we were able to measure \bafe\ in each globular cluster, we measure a $\sigma_{\text{sys}}=0.11$, 0.07, and 0.09\,dex for NGC~2419, M68, and M92, respectively. The abundances can be seen in Figure \ref{M92}, and they are included in Table~\ref{tab:GCcatalog}. The error distribution when including the measured systematic errors are shown in Figure \ref{M92_sys_hist}. 

\begin{equation}
\label{eq:intrinsic_error}
\text{stddev}\left(\frac{\text{\bafe}-\langle\text{\bafe}\rangle}{\sqrt{\sigma_{\text{stat}}^2+\sigma_{\text{sys}}^2}}\right) = 1
\end{equation}

\renewcommand{\thetable}{\arabic{table}}
\begin{deluxetable*}{llcccccccc}
\tablecolumns{10}
\tablecaption{DEIMOS Multi-Element Abundances Catalog of Globular Cluster Stars\label{tab:GCcatalog}}
\tablehead{\colhead{dSph} & \colhead{Name} & \colhead{RA} & \colhead{Dec} & \colhead{\teff} & \colhead{\logg} & \colhead{$\xi$} & \colhead{\feh} & \colhead{\afe} & \colhead{\bafe} \\
\colhead{ } & \colhead{ } & \colhead{ } & \colhead{ } & \colhead{(K)} & \colhead{($g$ in cm\,s$^{-2}$)} & \colhead{(km\,s$^{-1}$)} & \colhead{(dex)} & \colhead{(dex)} & \colhead{(dex)}}
\startdata
M92 & Stet-M92-S1081 & 17h16m59.472 & +43d07m09.58 & $4971\pm23$ & 2.06 & 1.66 & $-2.45\pm0.10$ & $+0.31\pm0.09$ & $+0.03\pm0.17$ \\
M92 & Stet-M92-S1377 & 17h17m02.346 & +43d06m56.47 & $5429\pm34$ & 3.25 & 1.37 & $-2.35\pm0.11$ & $+0.23\pm0.12$ & $-0.33\pm0.16$ \\
M92 & Stet-M92-S1416 & 17h17m03.069 & +43d06m36.66 & $4983\pm20$ & 2.06 & 1.65 & $-2.37\pm0.10$ & $+0.25\pm0.09$ & $-0.11\pm0.15$ \\
M92 & Stet-M92-S1622 & 17h17m02.321 & +43d09m31.88 & $4781\pm30$ & 1.69 & 1.74 & $-2.35\pm0.10$ & $+0.24\pm0.09$ & $-0.16\pm0.14$ \\
M92 & Stet-M92-S1687 & 17h17m04.879 & +43d07m32.34 & $5431\pm25$ & 2.92 & 1.45 & $-2.27\pm0.10$ & $+0.33\pm0.10$ & $-0.30\pm0.18$ \\
M92 & Stet-M92-S2476 & 17h17m07.992 & +43d11m26.82 & $5162\pm26$ & 2.48 & 1.56 & $-2.49\pm0.10$ & $+0.36\pm0.09$ & $-0.17\pm0.12$ \\
M92 & Stet-M92-S2492 & 17h17m11.893 & +43d07m28.93 & $5721\pm29$ & 3.52 & 1.31 & $-2.19\pm0.11$ & $+0.22\pm0.12$ & $-0.16\pm0.18$ \\
M92 & Stet-M92-S2732 & 17h17m11.044 & +43d10m34.69 & $5428\pm27$ & 3.33 & 1.35 & $-2.33\pm0.11$ & $+0.04\pm0.13$ & $-0.30\pm0.15$ \\
M92 & Stet-M92-S2736 & 17h17m11.200 & +43d10m28.52 & $5053\pm23$ & 2.15 & 1.63 & $-2.47\pm0.10$ & $+0.31\pm0.09$ & $+0.20\pm0.13$ \\
M92 & Stet-M92-S2977 & 17h17m12.782 & +43d11m18.59 & $5495\pm29$ & 3.36 & 1.35 & $-2.27\pm0.11$ & $+0.27\pm0.12$ & $-0.09\pm0.17$ \\
\enddata
\tablecomments{Table~\ref{tab:GCcatalog} is published in its entirety in the machine-readable format. A portion is shown here for guidance regarding its form and content. The errors reported here already include the systematic errors for \feh, \afe, and \bafe\ found in Table~\ref{tab:syserr}. Stars are only included if the errors for \feh, \afe, and \bafe\ are less than 0.28\,dex.}
\end{deluxetable*}

\subsection{Barium Abundance Error Floor} 

By comparing our \bafe\ measurements to HRS measurements in the literature, we measure $\sigma_{sys}=0.23$. By forcing the \bafe\ measurements of $\approx30$ stars in NGC~2419, M68, and M92 to have no intrinsic dispersion, we measure $\sigma_{sys}=0.11$, 0.07, and 0.09, respectively. There is a clear discrepancy between the systematic error measured from the HRS comparison and the globular clusters.  The HRS comparison relies on a heterogeneous collection of literature sources, with different spectrographs, line lists, and analysis codes.  Some or most of the systematic error determined from HRS comes from this heterogeneity.  On the other hand, the globular cluster analysis is internal to our own homogeneous study. Therefore, we set our systematic error at 0.1\,dex for \bafe. For context, the statistical errors of our measurements range from 0.1 to 0.28\,dex with an average of 0.19\,dex. All of the abundance error floors used in our catalog can be seen in Table~\ref{tab:syserr}.

\begin{deluxetable}{lc}
\tablecolumns{2}
\tablecaption{Abundance Error Floor\label{tab:syserr}}
\tablehead{\colhead{Abundance} & \colhead{Error Floor}} 
\startdata
\feh & 0.101\\
\afe & 0.084\\
\bafe & 0.100
\enddata
\tablecomments{These systematic errors are included in the errors given in Table \ref{tab:catalog}.}
\end{deluxetable}

\section{Discussion}
\label{sec:discussion}

\begin{figure*}
\centering
\includegraphics[width=\linewidth]{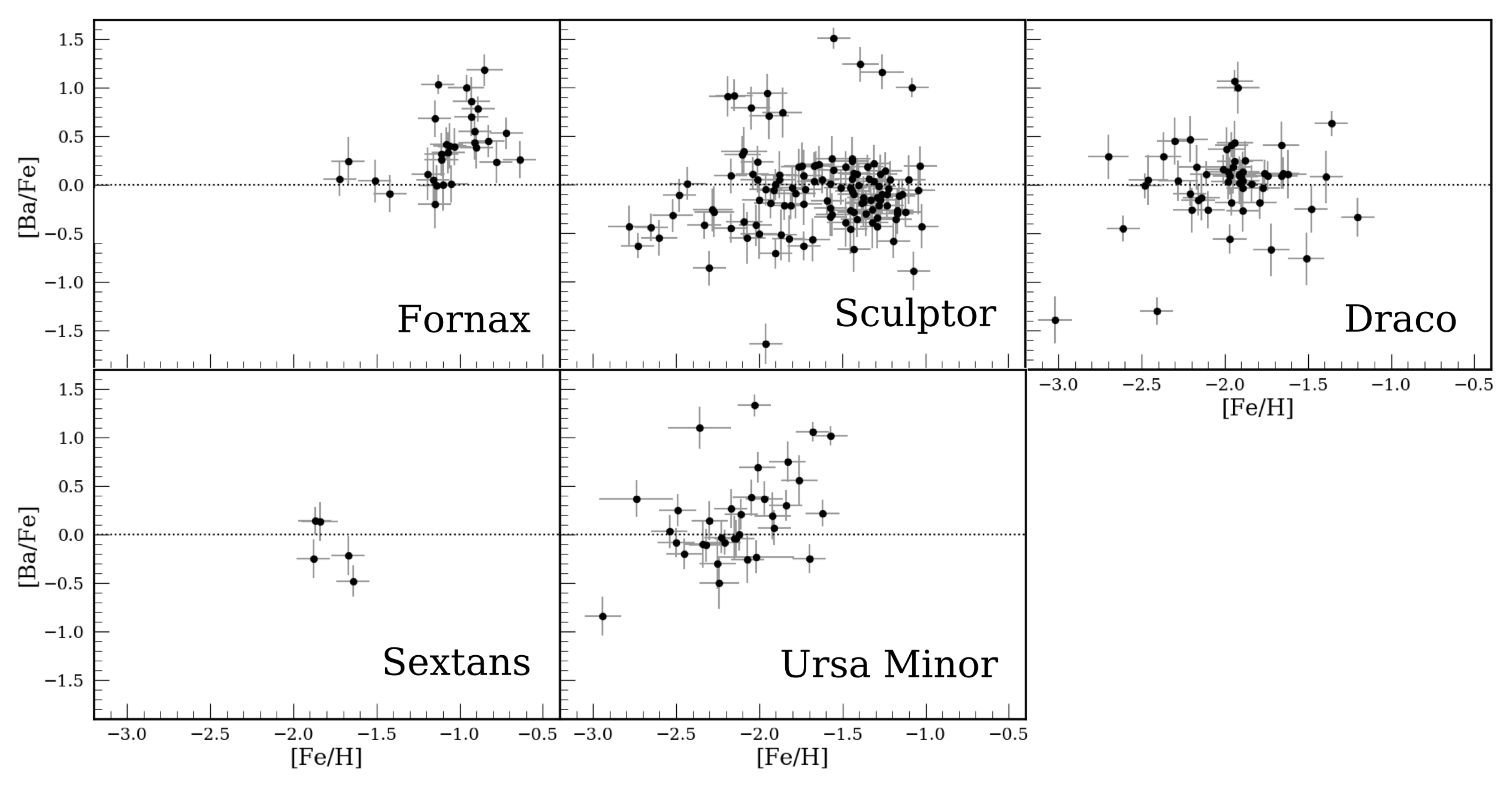}
\caption{Barium abundance measurements in Fornax, Sculptor, Draco, Sextans, and Ursa Minor dwarf galaxies. This is the largest sample of barium measurements in dwarf galaxies ever assembled with a total of 243 stars. Our discussion focuses on the abundance trends seen for each galaxy. In this discussion, we neglect the high \bafe\ outliers that are most clearly seen in Sculptor. These stars are likely to have been enriched by an AGB companion.
\label{dsph_barium}
}
\end{figure*}

We can observe neutron-capture abundance trends for several galaxies for the first time, because we have the largest sample of barium abundance measurements in dwarf galaxies ever assembled. We measured barium abundances in $\approx$250 stars with more than 30 red giant branch stars in each of Draco, Sculptor, Fornax, and Ursa Minor, in addition to five stars in Sextans (Figure \ref{dsph_barium}). This catalog of abundances increases the number of stars with barium measurements in these galaxies substantially, which can be seen by comparing the number of stars with published \bafe\ currently found in the literature ($N_{\text{LIT}}$) with the number of stars in our catalog ($N$) given in Table \ref{tab:properties}.

\begin{deluxetable*}{lcccccccccc}
\tablecolumns{11}
\tablecaption{Properties of Dwarf Galaxies\label{tab:properties}}
\tablehead{\colhead{Galaxy} &  \colhead{\phantom{\tablenotemark{a}}$D$\tablenotemark{a}} & \colhead{\phantom{\tablenotemark{b}}\mstar\tablenotemark{b}} & \colhead{\phantom{\tablenotemark{c}}$\langle\tau_{\text{SF}}\rangle$\tablenotemark{c}} & \colhead{\phantom{\tablenotemark{d}}$\tau_{90}$\tablenotemark{d}} & \colhead{\phantom{\tablenotemark{e}}\baeu$_{\text{med}}$\tablenotemark{e}} & \colhead{\phantom{\tablenotemark{f}}$f_r$\tablenotemark{f}} &\colhead{\phantom{\tablenotemark{g}}$\langle$\feh$\rangle$\tablenotemark{g}} &\colhead{\phantom{\tablenotemark{h}}N\tablenotemark{h}}&\colhead{\phantom{\tablenotemark{i}}$\langle$\bafe$\rangle$\tablenotemark{i}} &\colhead{\phantom{\tablenotemark{j}}N$_{\text{LIT}}$\tablenotemark{j}} 
\\
\colhead{ } & \colhead{(kpc)} & \colhead{($10^5$\,\msun)} & \colhead{(Gyr)} & \colhead{(Gyr)} & \colhead{(dex)} & \colhead{(\%)} & \colhead{(dex)} & \colhead{} & \colhead{(dex)} &  \colhead{}}  
\startdata
Fornax     & $147\pm9$ & $190\pm50$  & 7.4  & 2.3 & $-0.19$ & 33\% & $-0.99\pm0.01$ & 30 & $0.50\pm0.03$ & 106\\
Sculptor   & $85\pm4$  & $12\pm5$    & 12.6 & 10.8 & $-0.27$ & 40\% & $-1.68\pm0.01$ & 119 & $-0.02\pm0.02$ & 16 \\
Draco      & $75\pm5$  & $9.1\pm1.4$ & 10.9 & 10.2 & $-0.43$ & 58\% & $-1.93\pm0.01$ & 55 & $0.04\pm0.02$ & 24 \\ 
Sextans    & $85\pm3$  & $8.5\pm2.4$ & 12.0 & 12.9 & \nodata & \nodata & $-1.93\pm0.01$ & 5 & $-0.12\pm0.08$ & 12\\
Ursa Minor & $75\pm3$  & $5.6\pm1.7$ & 12.0 & 9.0 & $-0.57$ & 80\% & $-2.13\pm0.01$ & 34 & $0.33\pm0.03$ & 20
\enddata
\tablenotetext{a}{Distance from the MW \citep[][and references therein]{Kirby14}.}
\tablenotetext{b}{Stellar masses from \citet{Woo08}.}
\tablenotetext{c}{Mean mass-weighted stellar age from \citet{Orban08}.}
\tablenotetext{d}{Age at which 90\% of the stellar mass formed from \citeauthor{Weisz14}'s (\citeyear{Weisz14}) cumulative star formation histories. Sextans is not included in \citet{Weisz14}; therefore, we use the end of star formation from \citet{Bettinelli18}. This is used as an indication of the duration of star formation.}
\tablenotetext{e}{Median \baeu\ measurements from the literature. See Figure~\ref{BaEu} for detailed references.} 
\tablenotetext{f}{\baeu$_{\text{med}}$ converted to the percentage of \rproc\ contribution using $r$- and \sproc\ abundances published by \citet{Simmerer04}, as described in Equation~\ref{eq:rprocfrac}.}
\tablenotetext{g}{Mean \feh\ from \citet{Kirby11b} is weighted by the statistical uncertainty of each star.}
\tablenotetext{h}{Number of stars that are members and have \bafe\ measured in our catalog with errors less than 0.30\,dex.}
\tablenotetext{i}{Mean \bafe\ of our catalog is weighted by the statistical uncertainty of each star.}
\tablenotetext{j}{Number of unique stars that are members and have \bafe\ measurements published in various papers. The SAGA Database \citep{Suda17} was primarily used to compile the literature values with recent papers added separately. This total does not include stars with upper/lower limits. See Figure~\ref{mgba_trend} for specific references, with the exception of Fornax where we've also included stars from \citet{Letarte10} in the total of N$_{\text{LIT}}$.}
\end{deluxetable*}

Before we discuss the abundance trends in these galaxies, it is important to discuss the very high \bafe\ outliers that make up $\lesssim 10\%$ our sample. They are particularly obvious in Sculptor (10 stars with $\text{\bafe}>0.5$), but likely contaminate the other dwarf galaxies as well. We believe that many of these barium-rich outliers are stars that have been in a binary with an AGB star at some point in their evolution. That AGB companion transferred \sproc\ rich material onto the surface of the secondary star. The primary star since evolved into a white dwarf, leaving us to measure the polluted secondary red giant. Some literature studies have found similarly barium-rich stars ($\text{\bafe}\approx0.5$) and were able to use abundances of additional elements to confirm that they were enriched by an AGB companion \citep[e.g.,][]{Honda11}. Additional elemental abundances would be required to prove that these outliers in our sample have been polluted by AGB stars, which is beyond the scope of this paper. If these stars have been polluted, they do not represent the chemical composition of the gas from which they were born and should therefore be ignored when discussing abundance trends. However, we leave these stars in our sample because we do not have confirmation (from light elements or radial velocity variations) that these stars were enriched by an AGB companion. The outliers are relatively rare, so they do not significantly bias our results.

\subsection{Why AGB Stars are Not the Dominant Source of Barium Enrichment at Early Times}

In Section~\ref{sec:potential_sources} we isolated the potential large contributors of barium to AGB stars for the \sproc, and either a rare type of CCSNe (e.g., MRSNe) or NSMs for the \rproc. We now consider the \sproc\ source of barium, AGB stars, and discuss why they are not expected to be the dominant source of barium at early times.

In the solar system, barium is primarily produced by the \sproc\ \citep[85\%,][]{Simmerer04}, which mainly occurs in AGB stars. However, each AGB star produces a small amount of barium \citep[$\lesssim 10^{-6}$\,\msun\ of barium from AGB stars with $\text{\feh}\approx -0.7$;][]{Karakas18}. In old, metal-poor stellar populations there has not been sufficient time for AGB stars to significantly contribute neutron-capture elements, and barium is instead an indicator of the \rproc\ \citep[e.g.,][]{Ji16}. Combining the knowledge currently available in the literature of the SFHs and \baeu\ abundances for these dwarf galaxies confirms that AGB stars do not significantly contribute barium at early times (\feh\,$\lesssim -1.6$), and an \rproc\ site is responsible for the majority of barium enrichment observed.

If the SFH is short, we will not see AGB stars contribute significantly because stars stopped forming before the bulk of the low-mass AGB stars evolved to the point of ejecting barium. Table \ref{tab:properties} shows two independent tracers of the SFH: the average ages of the stars in the galaxies, $\langle\tau_{\text{SF}}\rangle$, and the time at which 90\% of the stars in the galaxy have been formed, $\tau_{90}$. Both of these metrics confirm that although Fornax has had a comparably extended SFH, all of the other dwarf galaxies have short SFHs that could at most last 2--3\,Gyrs. From the SFHs we would expect AGB stars to have the largest impact in Fornax because it has more extended star formation than the other dwarf galaxies. The contribution from AGB stars would only dominate the \bafe\ trend in Sculptor, Ursa Minor, and Draco at late times (the iron-rich end), and therefore would not dominate barium production at low \feh.

\begin{figure}
\centering
\includegraphics[width=\linewidth]{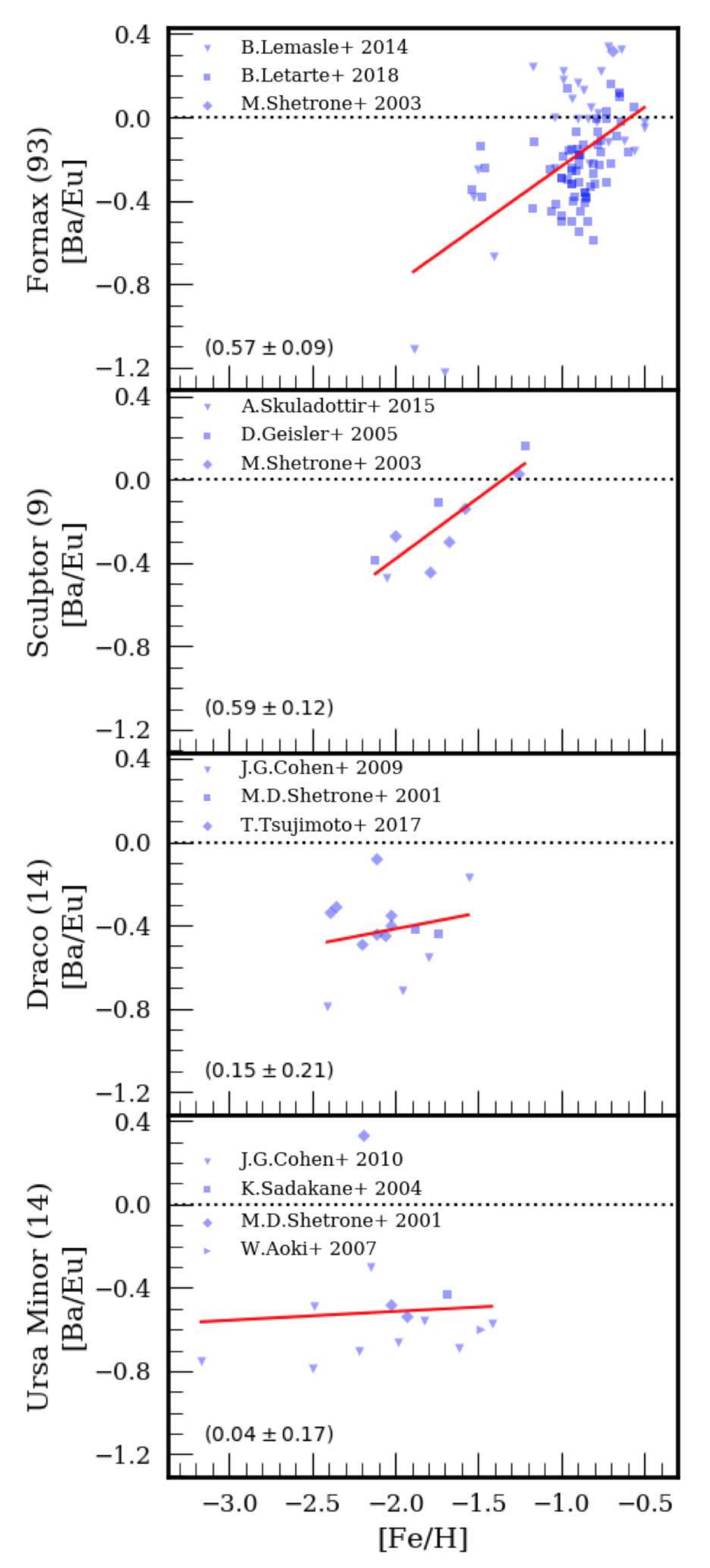}
\caption{Literature \baeu\ measurements for each galaxy. Additional \baeu\ measurements for Sculptor by Hill et al. (in prep) show a lower \baeu\ trend than shown in this plot (see Figure 14 of \citealt{Tolstoy09}). Sextans is not included because no \baeu\ detections exist. A line ($\text{\baeu} = A\times\text{\feh} + B$) is fit to each galaxy's abundances to determine the fraction of barium contributed via the \rproc. The slope ($A$) of the line for each galaxy is printed on the bottom left. 
\label{BaEu}}
\end{figure}

These conclusions are further confirmed by studying the \baeu\ measurements found in the literature (see Figure~\ref{BaEu}).  We convert these \baeu\ ratios to the percentage of neutron-capture elements contributed by the \rproc\ ($f_r$) using the pure $r$- and \sproc\ abundances reported by \citet{Simmerer04}. Equation~\ref{eq:rprocfrac} describes this conversion. In this equation $N_{s/r} = 10^{\log\epsilon_{s/r}}$, where $\log\epsilon_{s/r}$ is from Table~10 in \citet{Simmerer04}, and $f_r$ was restricted to remain between zero and one.  The solar ratio $({\rm Ba/Eu})_\odot$ is given by Table~\ref{sun}.

\begin{equation}
\label{eq:rprocfrac}
f_r(\text{\baeu}) = \displaystyle\frac{\frac{N_{\text{Eu}-s}}{N_{\text{Ba}-s}}-10^{-(\text{\baeu}+\text{(Ba/Eu)}_\odot)}}{\frac{N_{\text{Eu}-s}}{N_{\text{Ba}-s}}-\frac{N_{\text{Eu}-r}}{N_{\text{Ba}-r}}}
\end{equation}

\noindent The median \rproc\ contribution in each galaxy is included in Table \ref{tab:properties}. The lowest \rproc\ (i.e., the highest \sproc) contributions occur in galaxies with the longest SFHs. Although we do see \sproc\ (AGB) contribution dominate the barium contribution at higher \feh\ for Fornax and Sculptor, the \baeu\ trends in most of these galaxies have \baeu\,$\lesssim-0.4$ for \feh\,$\lesssim-1.6$. Sculptor is the exception to this statement (see Figure~\ref{BaEu}). However, extensive \baeu\ measurements by Hill et al. (in prep) in Sculptor are lower than the measurements by \citet{Shetrone03} and \citet{Geisler05}.  The newer measurements do satisfy the criterion of \baeu\,$\lesssim-0.4$ for \feh\,$\lesssim-1.6$ (these measurements are shown in Figure 14 of \citealt{Tolstoy09}). A \baeu\ value of $-0.4$ is equivalent to 54\% of barium being contributed by the \rproc. Therefore, AGB stars can not be the dominant source of barium at early times (\feh\,$\lesssim-1.6$).
\\
\subsection{Isolating the R-Process Contribution of Barium}

Although AGB stars are not the dominant source of barium at early times, we still need to remove their contribution to clearly study the \rproc. To isolate the barium contributed by the \rproc\ exclusively, we utilize the \baeu\ measurements found in the literature to subtract the \sproc\ contribution. We first fit a line to the \baeu\ and \feh\ measurements found in the literature (see Figure~\ref{BaEu}). This line can be converted (via Equation~\ref{eq:rprocfrac}) to the fraction of barium coming from the \rproc\ ($f_r$). We can now convert \bafe\ from our catalog (Figure~\ref{dsph_barium}) and the literature to include only the \rproc\ component by 
\bafe$_r = \text{\bafe} + \log (f_r)$. These results are displayed in the right panels of Figure~\ref{mgba_trend}. Now that we have isolated the \rproc\ component of barium, we can discuss the \rproc\ barium trend that we observe.

\subsection{New Critical Piece of Evidence of the Dominant R-Process Origin}

\begin{figure*}
\centering
\includegraphics[width=\linewidth]{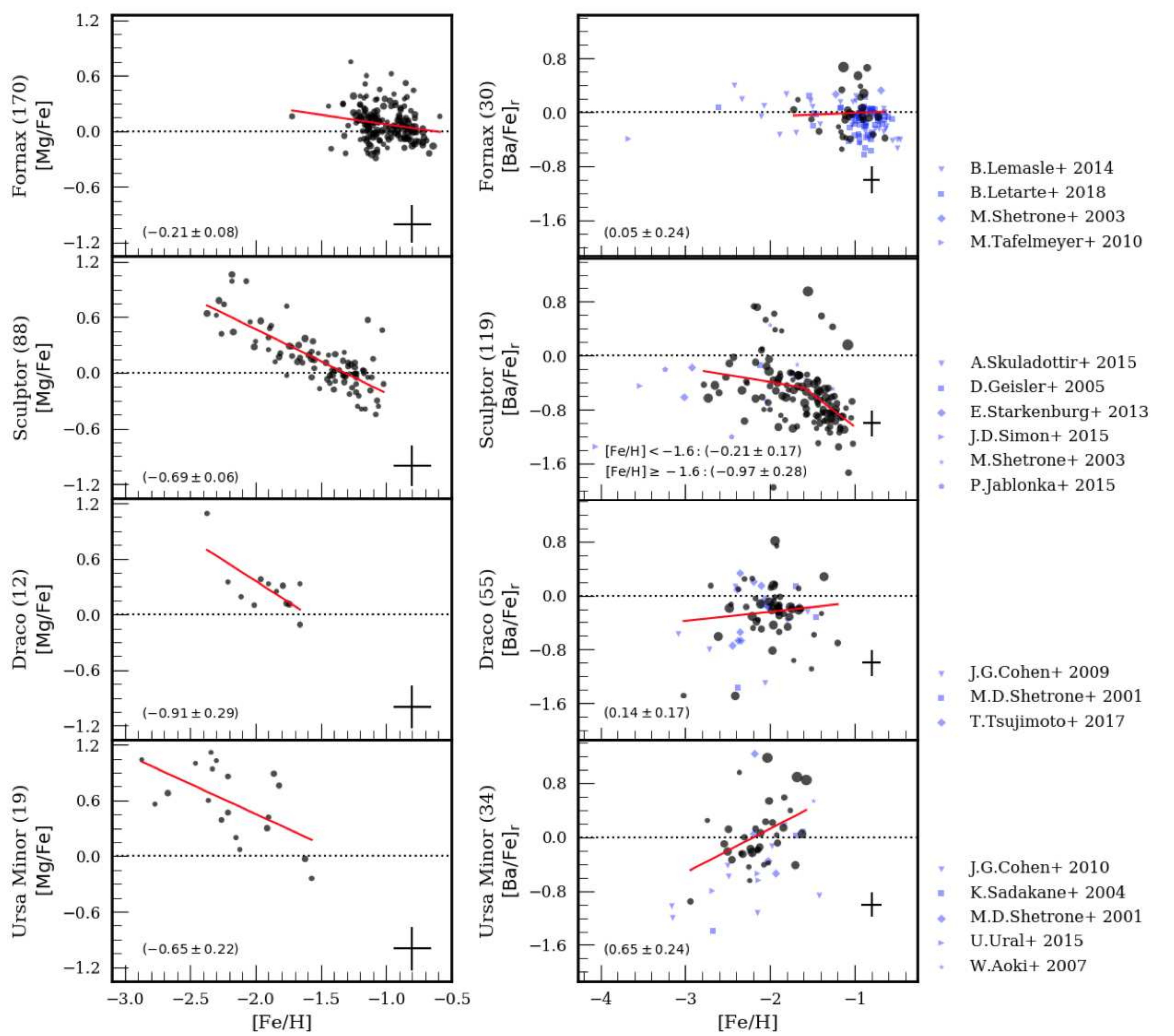}
\caption{Comparison of the trend of \mgfe\ \citep{Kirby10} and \bafe\ created by the \rproc\ (\bafer) as a function of \feh\ for each galaxy. The current published \bafe\ abundances (adjusted to include only the \rproc\ component) are shown in blue with the corresponding reference listed on the side. Overplotted in black are our \bafer\ measurements. The sizes of the dots are inversely proportional to the errors, and the average errors for the stars plotted in each sub-figure are shown in the bottom right. The stars in black have abundance errors less than 0.28\,dex and the total number plotted is listed in parentheses. A linear fit is shown in red, and the slope of this line is printed in the bottom left of each panel. The main conclusion is that \mgfe\ decreases as a function of \feh\ while \bafe\ has a significantly more positive slope for low metallicities. This indicates that barium is contributed on a more delayed timescale than CCSNe.
\label{mgba_trend}}
\end{figure*}

The key result of this paper is shown in Figure~\ref{mgba_trend} where we compare \afe\ \citep[specifically \mgfe,][]{Kirby10} as a function of \feh\ to the \rproc\ component of \bafe\ (\bafer) as a function of \feh. Barium results from both this catalog (black circles) and the literature values (blue) are displayed to utilize all the available information. Linear fits to \citeauthor{Kirby10}'s and our catalog are shown in red with the slope of this fit printed for each galaxy. The trend is that \mgfe\ decreases with iron abundance because CCSNe contribute $\alpha$ elements before SNe Ia contribute large quantities of iron. However the barium abundances have slopes that are significantly more positive (even when accounting for the uncertainties of the slopes) for \feh\,$\lesssim-1.6$. This can be seen in Sculptor, Draco, and Ursa Minor. When possible, we allowed the slope to differ for \feh\ greater than and less than $-1.6$. There were only enough stars above and below this \feh\ cutoff in Sculptor to allow this broken slope. The \feh\ cutoff of $-1.6$ was chosen because visually the slope of \bafe\ vs. \feh\ changes in Sculptor at approximately this \feh\ abundance.

The fact that all of the galaxies in our sample simultaneously have a negative \mgfe\ vs. \feh\ slope and a significantly more positive \bafer\ vs. \feh\ slope for \feh\,$\lesssim-1.6$ leads to the conclusion that \rproc\ barium is delayed relative to magnesium at early times. Therefore, Ba cannot come from the same source (CCSNe) as Mg.

\nocite{Lemasle14,Shetrone03,Tafelmeyer10,Geisler05,Starkenburg13,Simon15,Jablonka15,Cohen09,Shetrone01,Tsujimoto17,Aoki09,Cohen10,Sadakane04,Ural15,Aoki07,Letarte18,Skuladottir15}

The discrepant slopes of \mgfe\ and \bafer\ vs. \feh\ are a powerful diagnostic feature of the \rproc\ origin for a few different reasons. First, we are comparing consistent samples with the same techniques for these abundance measurements. Second, we are comparing abundance trends for the same galaxies. Therefore, both \mgfe\ and \bafer\ are subject to the same SFH, gas inflow/outflow history, and \feh\ trend with time. This removes the dependence of our conclusion on a galactic chemical evolution model and its myriad assumptions. Third, we see the discrepant slopes in multiple dwarf galaxies. This proves that we are seeing a global characteristic of the \rproc.

Our observation of a significantly more positive slope of \bafer\ vs. \feh\ than \afe for low iron abundances (\feh\,$\lesssim-1.6$) in several dwarf galaxies, requires that the timescale for barium is substantially more delayed than the timescale for magnesium. We now consider the possible origins for barium and which source would cause the observed trends. 

\subsection{NSMs Could Be the Dominant Source of Barium in Dwarf Galaxies}

We have concluded that dwarf galaxies are dominated by a \rproc\ source with a timescale more delayed than CCSNe at early times (\feh\,$\lesssim-1.6$). Compared to ultra-faint dwarfs \citep{Ji16,Hansen17}, our sample of classical dwarf galaxies is able to probe the characteristics of an ensemble of \rproc\ enrichment rather than isolated events. This evidence includes information about the timescale of enrichment, which enables us to distinguish whether MRSNe or NSMs are the dominant source of \rproc\ enrichment in dwarf galaxies at early times.

First we will consider MRSNe. All CCSNe have a short lifetime, so they would cause the same negative trend that we see in the $\alpha$ elements, which are also released by CCSNe.  Even if the $r$-process arises from very rare types of CCSN, like MRSNe, we would expect an overall negative---and perhaps jagged---slope in [$r$/Fe] vs.\ [Fe/H]\@. Because we observe a significantly more positive \bafer\ vs. \feh\ slope, we can rule out MRSNe as the dominant source of barium in these dwarf galaxies.  Common envelope jets SNe have a slightly longer delay time than CCSNe, but the delay time is still set by the lifetime of the secondary star in the binary system.  Because both stars in this scenario are massive, their delay times would be much closer to CCSNe than NSMs, whose delay times are set by their protracted orbital decay.

Instead, we need a delayed barium enrichment to create a more positive \bafer\ vs. \feh\ slope when compared to the \mgfe\ vs. \feh\ slope. A flat \bafer\ vs. \feh\ trend would indicate that the \rproc\ enrichment needs to occur on a timescale similar to SNe~Ia because barium would need to be ejected into the ISM while SNe~Ia are ejecting large amounts of iron. Although we see a clear increasing \bafer\ vs. \feh\ trend in Ursa Minor and a flat or slightly rising trend in Draco, Sculptor has a slightly negative trend for $\text{\feh}\lesssim-1.6$. It is therefore ambiguous when the \rproc\ enrichment needs to occur compared to SNe~Ia timescales. However, we can definitively say that the timescale needs to be delayed compared to CCSNe in order to create the significantly more positive \bafer\ vs. \feh\ slope. Based on our observations, NSMs are the most viable source of barium enrichment in dwarf galaxies at early times. Simulations and the LIGO observation support NSMs producing \rproc\ elements \citep[e.g.,][]{Goriely11,Abbott17,Cote17}.

In conclusion, our observations are matched by a source that releases barium on a timescale more delayed than CCSNe at early times. Of the sources suggested so far, neutron star mergers are the only source that satisfies this condition.

\subsection{Implications}

The early chemical evolution of dwarf galaxies can be used to constrain the yield and/or rate of NSMs, which we will address with a galactic chemical evolution model in a future paper. We have concluded that \rproc\ enrichment is dominated by NSMs in the early evolution of dwarf galaxies. It is tempting to extrapolate this conclusion and apply this directly to comment on the dominant site of \rproc\ enrichment in later evolution of dwarf galaxies and larger galaxies (such as the Milky Way). 

However, we exhort readers to extrapolate to high \feh\ with caution. The rate of NSMs that have ejected material retained in the galaxy (and the yield or amount of ejected material that is retained per NSM) may depend on the age of the binary neutron star system and/or mass of the galaxy. To discuss the \rproc\ trends seen in the literature at higher iron abundances, we turn to \eufe\ as an indicator of the \rproc, because the \sproc\ is increasingly important in barium productions at these times. For example, Hill et al. (in prep) find a declining \eufe\ vs. \feh\ trend (see Figure 13 in \citealt{Tolstoy09}) in Sculptor at $\text{\feh}\gtrsim-1.6$. Comparisons to the Milky Way are more challenging due to the drastically different mass and merger history. Milky Way halo stars are largely a compilation of stars stripped from dwarf galaxies. It is therefore unsurprising that there is a large spread in \eufe. In Milky Way disk stars (\feh\,$\gtrsim-1.0$) a declining \eufe\ vs. \feh\ trend is also observed \citep{Battistini16}.  The complexity of the Milky Way relative to dwarf galaxies has led to varying conclusions as to whether the origin of the $r$-process is NSMs, MRSNe, or both \citep[e.g.,][]{Cescutti15,Wehmeyer15}.

This transition from positive [\rproc/Fe] vs. \feh\ trends at low metallicities to a declining trend at higher metallicities presents a puzzle. Although some attribute this to MRSNe \citep{Tsujimoto15_mrsn}, another possible explanation is that NSM natal kicks cause the effective NSM rate (rate of NSMs contributing enriched material to the ISM) to decrease significantly at later times as the NSMs occur far from the galaxy \citep[e.g.,][]{Willems04,Bramante16,Beniamini16a,Safarzadeh17}. It is important to note that the velocities of these NSM natal kicks are still unknown, and it is possible that they are large enough to remove the neutron star binary from the galaxy before a NSM can occur. Moderate NSM natal kicks that allow early NSMs to occur in the galaxy and late NSMs occur outside the galaxy would be consistent with our \bafer\ measurements, specifically the significant decrease in \bafer\ vs. \feh\ slope seen in Sculptor above and below $\text{\feh}=-1.6$). Additional detections of \eufe\ with \feh\,$\lesssim-1.6$ and modeling of NSM natal kicks would be able to confirm the plausibility of this explanation.
\\
\section{Summary}
Here we highlight the main conclusions of this paper.
\begin{itemize}
    \item We have confirmed medium-resolution spectroscopy as a reliable method of measuring barium. 
    \item We have obtained the largest sample of barium abundances in dwarf galaxies.
    \item We have discovered that the majority of barium in dwarf galaxies is created by a delayed \rproc\ source at early times.
    \item We conclude that neutron star mergers are the most likely source of barium enrichment in dwarf galaxies at early times.
\end{itemize}
In a subsequent paper we will use galactic chemical evolution models to  derive qualitative conclusions concerning the sources of barium enrichment in dwarf galaxies. Specifically, we will constrain the NSM yields/rates needed to match our observations.

\acknowledgements
This material is based upon work supported by the National Science Foundation Graduate Research Fellowship under Grant No. DGE‐1745301 and the National Science Foundation under Grant No.\ AST-1614081.  We thank Alexander Ji for insightful conversation.

\vspace{5mm}
\facility{Keck:II (DEIMOS)}

\software{MOOG \citep{Sneden73}, spec2d pipeline \citep{Cooper12,Newman13}, scipy \citep{Jones01}}

\bibliographystyle{aasjournal}
\bibliography{abundance}

\begin{thebibliography}{}
\expandafter\ifx\csname natexlab\endcsname\relax\def\natexlab#1{#1}\fi

\bibitem[{{Abbott} {et~al.}(2017){Abbott}, {Abbott}, {Abbott}, {Acernese},
  {Ackley}, {Adams}, {Adams}, {Addesso}, {Adhikari}, {Adya}, \&
  et~al.}]{Abbott17}
{Abbott}, B.~P., {Abbott}, R., {Abbott}, T.~D., {et~al.} 2017, Physical Review
  Letters, 119, 161101

\bibitem[{{Anders} \& {Grevesse}(1989)}]{Anders89}
{Anders}, E., \& {Grevesse}, N. 1989, \gca, 53, 197

\bibitem[{{Andrievsky} {et~al.}(2017){Andrievsky}, {Korotin}, {Hill}, \&
  {Zhukova}}]{Andrievsky17}
{Andrievsky}, S.~M., {Korotin}, S.~A., {Hill}, V., \& {Zhukova}, A.~V. 2017,
  arXiv:1710.04930

\bibitem[{{Andrievsky} {et~al.}(2009){Andrievsky}, {Spite}, {Korotin}, {Spite},
  {Fran{\c c}ois}, {Bonifacio}, {Cayrel}, \& {Hill}}]{Andrievsky09}
{Andrievsky}, S.~M., {Spite}, M., {Korotin}, S.~A., {et~al.} 2009, \aap, 494,
  1083

\bibitem[{{Aoki} {et~al.}(2009{\natexlab{a}}){Aoki}, {Barklem}, {Beers},
  {Christlieb}, {Inoue}, {Garc{\'\i}a P{\'e}rez}, {Norris}, \&
  {Carollo}}]{Aoki09_synth}
{Aoki}, W., {Barklem}, P.~S., {Beers}, T.~C., {et~al.} 2009{\natexlab{a}},
  \apj, 698, 1803

\bibitem[{{Aoki} {et~al.}(2007){Aoki}, {Honda}, {Sadakane}, \&
  {Arimoto}}]{Aoki07}
{Aoki}, W., {Honda}, S., {Sadakane}, K., \& {Arimoto}, N. 2007, \pasj, 59, L15

\bibitem[{{Aoki} {et~al.}(2009{\natexlab{b}}){Aoki}, {Arimoto}, {Sadakane},
  {Tolstoy}, {Battaglia}, {Jablonka}, {Shetrone}, {Letarte}, {Irwin}, {Hill},
  {Francois}, {Venn}, {Primas}, {Helmi}, {Kaufer}, {Tafelmeyer}, {Szeifert}, \&
  {Babusiaux}}]{Aoki09}
{Aoki}, W., {Arimoto}, N., {Sadakane}, K., {et~al.} 2009{\natexlab{b}}, \aap,
  502, 569

\bibitem[{{Arnould} {et~al.}(2007){Arnould}, {Goriely}, \&
  {Takahashi}}]{Arnould07}
{Arnould}, M., {Goriely}, S., \& {Takahashi}, K. 2007, \physrep, 450, 97

\bibitem[{{Barack} {et~al.}(2018){Barack}, {Cardoso}, {Nissanke}, {Sotiriou},
  {Askar}, {Belczynski}, {Bertone}, {Bon}, {Blas}, {Brito}, {Bulik}, {Burrage},
  {Byrnes}, {Caprini}, {Chernyakova}, {Chrusciel}, {Colpi}, {Ferrari},
  {Gaggero}, {Gair}, {Garcia-Bellido}, {Hassan}, {Heisenberg}, {Hendry},
  {Heng}, {Herdeiro}, {Hinderer}, {Horesh}, {Kavanagh}, {Kocsis}, {Kramer}, {Le
  Tiec}, {Mingarelli}, {Nardini}, {Nelemans}, {Palenzuela}, {Pani}, {Perego},
  {Porter}, {Rossi}, {Schmidt}, {Sesana}, {Sperhake}, {Stamerra}, {Stein},
  {Tamanini}, {Tauris}, {Urena-Lopez}, {Vincent}, {Volonteri}, {Wardell},
  {Wex}, {Yagi}, {Abdelsalhin}, {Aloy}, {Amaro-Seoane}, {Annulli},
  {Arca-Sedda}, {Bah}, {Barausse}, {Barakovic}, {Benkel}, {Bennett}, {Bernard},
  {Bernuzzi}, {Berry}, {Berti}, {Bezares}, {Juan Blanco-Pillado},
  {Blazquez-Salcedo}, {Bonetti}, {Boskovic}, {Bosnjak}, {Bricman}, {Bruegmann},
  {Capelo}, {Carloni}, {Cerda-Duran}, {Charmousis}, {Chaty}, {Clerici},
  {Coates}, {Colleoni}, {Collodel}, {Compere}, {Cook}, {Cordero-Carrion},
  {Correia}, {de la Cruz-Dombriz}, {Czinner}, {Destounis}, {Dialektopoulos},
  {Doneva}, {Dotti}, {Drew}, {Eckner}, {Edholm}, {Emparan}, {Erdem},
  {Ferreira}, {Ferreira}, {Finch}, {Font}, {Franchini}, {Fransen}, {Gal'tsov},
  {Ganguly}, {Gerosa}, {Glampedakis}, {Gomboc}, {Goobar}, {Gualtieri},
  {Guendelman}, {Haardt}, {Harmark}, {Hejda}, {Hertog}, {Hopper}, {Husa},
  {Ihanec}, {Ikeda}, {Jaodand}, {Jetzer Xisco Jimenez-Forteza}, {Kamionkowski},
  {Kaplan}, {Kazantzidis}, {Kimura}, {Kobayashi}, {Kokkotas}, {Krolik}, {Kunz},
  {Lammerzahl}, {Lasky}, {Lemos}, {Levi Said}, {Liberati}, {Lopes}, {Luna},
  {Ma}, {Maggio}, {Martinez Montero}, {Maselli}, {Mayer}, {Mazumdar},
  {Messenger}, {Menard}, {Minamitsuji}, {Moore}, {Mota}, {Nampalliwar},
  {Nerozzi}, {Nichols}, {Nissimov}, {Obergaulinger}, {Obers}, {Oliveri},
  {Pappas}, {Pasic}, {Peiris}, {Petrushevska}, {Pollney}, {Pratten}, {Rakic},
  {Racz}, {Ramazanouglu}, {Ramos-Buades}, {Raposo}, {Rogatko}, {Rosinska},
  {Rosswog}, {Ruiz Morales}, {Sakellariadou}, {Sanchis-Gual}, {Sharan Salafia},
  {Sintes}, {Smole}, {Sopuerta}, {Souza-Lima}, {Stalevski}, {Stergioulas},
  {Stevens}, {Tamfal}, {Torres-Forne}, {Tsygankov}, {Unluturk}, {Valiante},
  {Velhinho}, {Verbin}, {Vercnocke}, {Vernieri}, {Vicente}, {Vitagliano},
  {Weltman}, {Whiting}, {Williamson}, {Witek}, {Wojnar}, {Yakut}, {Yan},
  {Yazadjiev}, {Zaharijas}, \& {Zilhao}}]{Barack18}
{Barack}, L., {Cardoso}, V., {Nissanke}, S., {et~al.} 2018, ArXiv e-prints,
  arXiv:1806.05195

\bibitem[{{Battistini} \& {Bensby}(2016)}]{Battistini16}
{Battistini}, C., \& {Bensby}, T. 2016, \aap, 586, A49

\bibitem[{{Bellazzini} {et~al.}(2002){Bellazzini}, {Ferraro}, {Origlia},
  {Pancino}, {Monaco}, \& {Oliva}}]{Bellazzini02}
{Bellazzini}, M., {Ferraro}, F.~R., {Origlia}, L., {et~al.} 2002, \aj, 124,
  3222

\bibitem[{{Beniamini} {et~al.}(2018){Beniamini}, {Dvorkin}, \&
  {Silk}}]{Beniamini18}
{Beniamini}, P., {Dvorkin}, I., \& {Silk}, J. 2018, \mnras, 478, 1994

\bibitem[{{Beniamini} {et~al.}(2016{\natexlab{a}}){Beniamini}, {Hotokezaka}, \&
  {Piran}}]{Beniamini16a}
{Beniamini}, P., {Hotokezaka}, K., \& {Piran}, T. 2016{\natexlab{a}}, \apjl,
  829, L13

\bibitem[{{Beniamini} {et~al.}(2016{\natexlab{b}}){Beniamini}, {Hotokezaka}, \&
  {Piran}}]{Beniamini16b}
---. 2016{\natexlab{b}}, \apj, 832, 149

\bibitem[{{Bensby} {et~al.}(2014){Bensby}, {Feltzing}, \& {Oey}}]{Bensby14}
{Bensby}, T., {Feltzing}, S., \& {Oey}, M.~S. 2014, \aap, 562, A71

\bibitem[{{Bettinelli} {et~al.}(2018){Bettinelli}, {Hidalgo}, {Cassisi},
  {Aparicio}, \& {Piotto}}]{Bettinelli18}
{Bettinelli}, M., {Hidalgo}, S.~L., {Cassisi}, S., {Aparicio}, A., \& {Piotto},
  G. 2018, \mnras, 476, 71

\bibitem[{{Bramante} \& {Linden}(2016)}]{Bramante16}
{Bramante}, J., \& {Linden}, T. 2016, \apj, 826, 57

\bibitem[{{Castelli} \& {Kurucz}(2004)}]{Castelli04}
{Castelli}, F., \& {Kurucz}, R.~L. 2004, ArXiv Astrophysics e-prints,
  astro-ph/0405087

\bibitem[{{Cescutti} {et~al.}(2015){Cescutti}, {Romano}, {Matteucci},
  {Chiappini}, \& {Hirschi}}]{Cescutti15}
{Cescutti}, G., {Romano}, D., {Matteucci}, F., {Chiappini}, C., \& {Hirschi},
  R. 2015, \aap, 577, A139

\bibitem[{{Cohen}(2011)}]{Cohen11}
{Cohen}, J.~G. 2011, \apjl, 740, L38

\bibitem[{{Cohen} \& {Huang}(2009)}]{Cohen09}
{Cohen}, J.~G., \& {Huang}, W. 2009, \apj, 701, 1053

\bibitem[{{Cohen} \& {Huang}(2010)}]{Cohen10}
---. 2010, \apj, 719, 931

\bibitem[{{Cohen} \& {Kirby}(2012)}]{Cohen12}
{Cohen}, J.~G., \& {Kirby}, E.~N. 2012, \apj, 760, 86

\bibitem[{{Cooper} {et~al.}(2012){Cooper}, {Newman}, {Davis}, {Finkbeiner}, \&
  {Gerke}}]{Cooper12}
{Cooper}, M.~C., {Newman}, J.~A., {Davis}, M., {Finkbeiner}, D.~P., \& {Gerke},
  B.~F. 2012, {spec2d: DEEP2 DEIMOS Spectral Pipeline}, Astrophysics Source
  Code Library, , , ascl:1203.003

\bibitem[{{C{\^o}t{\'e}} {et~al.}(2017){C{\^o}t{\'e}}, {Belczynski}, {Fryer},
  {Ritter}, {Paul}, {Wehmeyer}, \& {O'Shea}}]{Cote17}
{C{\^o}t{\'e}}, B., {Belczynski}, K., {Fryer}, C.~L., {et~al.} 2017, \apj, 836,
  230

\bibitem[{{Coulter} {et~al.}(2017){Coulter}, {Foley}, {Kilpatrick}, {Drout},
  {Piro}, {Shappee}, {Siebert}, {Simon}, {Ulloa}, {Kasen}, {Madore},
  {Murguia-Berthier}, {Pan}, {Prochaska}, {Ramirez-Ruiz}, {Rest}, \&
  {Rojas-Bravo}}]{Coulter17}
{Coulter}, D.~A., {Foley}, R.~J., {Kilpatrick}, C.~D., {et~al.} 2017, Science,
  358, 1556

\bibitem[{{De Cia} {et~al.}(2018){De Cia}, {Gal-Yam}, {Rubin}, {Leloudas},
  {Vreeswijk}, {Perley}, {Quimby}, {Yan}, {Sullivan}, {Fl{\"o}rs}, {Sollerman},
  {Bersier}, {Cenko}, {Gal-Yam}, {Maguire}, {Ofek}, {Prentice}, {Schulze},
  {Spyromilio}, {Valenti}, {Arcavi}, {Corsi}, {Howell}, {Mazzali}, {Kasliwal},
  {Taddia}, \& {Yaron}}]{DeCia18}
{De Cia}, A., {Gal-Yam}, A., {Rubin}, A., {et~al.} 2018, \apj, 860, 100

\bibitem[{{Escala} {et~al.}(2018){Escala}, {Wetzel}, {Kirby}, {Hopkins}, {Ma},
  {Wheeler}, {Kere{\v s}}, {Faucher-Gigu{\`e}re}, \& {Quataert}}]{Escala18}
{Escala}, I., {Wetzel}, A., {Kirby}, E.~N., {et~al.} 2018, \mnras, 474, 2194

\bibitem[{{Evans} {et~al.}(2017){Evans}, {Cenko}, {Kennea}, {Emery}, {Kuin},
  {Korobkin}, {Wollaeger}, {Fryer}, {Madsen}, {Harrison}, {Xu}, {Nakar},
  {Hotokezaka}, {Lien}, {Campana}, {Oates}, {Troja}, {Breeveld}, {Marshall},
  {Barthelmy}, {Beardmore}, {Burrows}, {Cusumano}, {D'A{\`i}}, {D'Avanzo},
  {D'Elia}, {de Pasquale}, {Even}, {Fontes}, {Forster}, {Garcia}, {Giommi},
  {Grefenstette}, {Gronwall}, {Hartmann}, {Heida}, {Hungerford}, {Kasliwal},
  {Krimm}, {Levan}, {Malesani}, {Melandri}, {Miyasaka}, {Nousek}, {O'Brien},
  {Osborne}, {Pagani}, {Page}, {Palmer}, {Perri}, {Pike}, {Racusin}, {Rosswog},
  {Siegel}, {Sakamoto}, {Sbarufatti}, {Tagliaferri}, {Tanvir}, \&
  {Tohuvavohu}}]{Evans17}
{Evans}, P.~A., {Cenko}, S.~B., {Kennea}, J.~A., {et~al.} 2017, Science, 358,
  1565

\bibitem[{{Faber} {et~al.}(2003){Faber}, {Phillips}, {Kibrick}, {Alcott},
  {Allen}, {Burrous}, {Cantrall}, {Clarke}, {Coil}, {Cowley}, {Davis}, {Deich},
  {Dietsch}, {Gilmore}, {Harper}, {Hilyard}, {Lewis}, {McVeigh}, {Newman},
  {Osborne}, {Schiavon}, {Stover}, {Tucker}, {Wallace}, {Wei}, {Wirth}, \&
  {Wright}}]{Faber03}
{Faber}, S.~M., {Phillips}, A.~C., {Kibrick}, R.~I., {et~al.} 2003, in
  \procspie, Vol. 4841, Instrument Design and Performance for Optical/Infrared
  Ground-based Telescopes, ed. M.~{Iye} \& A.~F.~M. {Moorwood}, 1657--1669

\bibitem[{{Foucart} {et~al.}(2015){Foucart}, {O'Connor}, {Roberts}, {Duez},
  {Haas}, {Kidder}, {Ott}, {Pfeiffer}, {Scheel}, \& {Szilagyi}}]{Foucart15}
{Foucart}, F., {O'Connor}, E., {Roberts}, L., {et~al.} 2015, \prd, 91, 124021

\bibitem[{{Fulbright}(2000)}]{Fulbright00}
{Fulbright}, J.~P. 2000, \aj, 120, 1841

\bibitem[{{Fulbright}(2002)}]{Fulbright02}
---. 2002, \aj, 123, 404

\bibitem[{{Geisler} {et~al.}(2005){Geisler}, {Smith}, {Wallerstein},
  {Gonzalez}, \& {Charbonnel}}]{Geisler05}
{Geisler}, D., {Smith}, V.~V., {Wallerstein}, G., {Gonzalez}, G., \&
  {Charbonnel}, C. 2005, \aj, 129, 1428

\bibitem[{{Gilmore} \& {Wyse}(1991)}]{Gilmore91}
{Gilmore}, G., \& {Wyse}, R.~F.~G. 1991, \apjl, 367, L55

\bibitem[{{Goriely} {et~al.}(2011){Goriely}, {Bauswein}, \&
  {Janka}}]{Goriely11}
{Goriely}, S., {Bauswein}, A., \& {Janka}, H.-T. 2011, \apjl, 738, L32

\bibitem[{{Grichener} \& {Soker}(2018)}]{Grichener18}
{Grichener}, A., \& {Soker}, N. 2018, ArXiv e-prints, arXiv:1810.03889

\bibitem[{{Gustafsson} {et~al.}(1975){Gustafsson}, {Bell}, {Eriksson}, \&
  {Nordlund}}]{Gustafsson75}
{Gustafsson}, B., {Bell}, R.~A., {Eriksson}, K., \& {Nordlund}, A. 1975, \aap,
  42, 407

\bibitem[{{Gustafsson} {et~al.}(2008){Gustafsson}, {Edvardsson}, {Eriksson},
  {J{\o}rgensen}, {Nordlund}, \& {Plez}}]{Gustafsson08}
{Gustafsson}, B., {Edvardsson}, B., {Eriksson}, K., {et~al.} 2008, \aap, 486,
  951

\bibitem[{{Gustafsson} {et~al.}(2003){Gustafsson}, {Edvardsson}, {Eriksson},
  {Mizuno-Wiedner}, {J{\o}rgensen}, \& {Plez}}]{Gustafsson03}
{Gustafsson}, B., {Edvardsson}, B., {Eriksson}, K., {et~al.} 2003, in
  Astronomical Society of the Pacific Conference Series, Vol. 288, Stellar
  Atmosphere Modeling, ed. I.~{Hubeny}, D.~{Mihalas}, \& K.~{Werner}, 331

\bibitem[{{Hansen} {et~al.}(2017){Hansen}, {Simon}, {Marshall}, {Li},
  {Carollo}, {DePoy}, {Nagasawa}, {Bernstein}, {Drlica-Wagner}, {Abdalla},
  {Allam}, {Annis}, {Bechtol}, {Benoit-L{\'e}vy}, {Brooks}, {Buckley-Geer},
  {Carnero Rosell}, {Carrasco Kind}, {Carretero}, {Cunha}, {da Costa}, {Desai},
  {Eifler}, {Fausti Neto}, {Flaugher}, {Frieman}, {Garc{\'{\i}}a-Bellido},
  {Gaztanaga}, {Gerdes}, {Gruen}, {Gruendl}, {Gschwend}, {Gutierrez}, {James},
  {Krause}, {Kuehn}, {Kuropatkin}, {Lahav}, {Miquel}, {Plazas}, {Romer},
  {Sanchez}, {Santiago}, {Scarpine}, {Smith}, {Soares-Santos}, {Sobreira},
  {Suchyta}, {Swanson}, {Tarle}, {Walker}, \& {DES Collaboration}}]{Hansen17}
{Hansen}, T.~T., {Simon}, J.~D., {Marshall}, J.~L., {et~al.} 2017, \apj, 838,
  44

\bibitem[{{Harris}(1996)}]{Harris96}
{Harris}, W.~E. 1996, \aj, 112, 1487

\bibitem[{{Hinkle} {et~al.}(2000){Hinkle}, {Wallace}, {Valenti}, \&
  {Harmer}}]{Hinkle00}
{Hinkle}, K., {Wallace}, L., {Valenti}, J., \& {Harmer}, D. 2000, {Visible and
  Near Infrared Atlas of the Arcturus Spectrum 3727-9300 A}

\bibitem[{{Honda} {et~al.}(2011){Honda}, {Aoki}, {Arimoto}, \&
  {Sadakane}}]{Honda11}
{Honda}, S., {Aoki}, W., {Arimoto}, N., \& {Sadakane}, K. 2011, \pasj, 63, 523

\bibitem[{{Iwamoto} {et~al.}(1999){Iwamoto}, {Brachwitz}, {Nomoto},
  {Kishimoto}, {Umeda}, {Hix}, \& {Thielemann}}]{Iwamoto99}
{Iwamoto}, K., {Brachwitz}, F., {Nomoto}, K., {et~al.} 1999, \apjs, 125, 439

\bibitem[{{Jablonka} {et~al.}(2015){Jablonka}, {North}, {Mashonkina}, {Hill},
  {Revaz}, {Shetrone}, {Starkenburg}, {Irwin}, {Tolstoy}, {Battaglia}, {Venn},
  {Helmi}, {Primas}, \& {Fran{\c c}ois}}]{Jablonka15}
{Jablonka}, P., {North}, P., {Mashonkina}, L., {et~al.} 2015, \aap, 583, A67

\bibitem[{{Ji} \& {Frebel}(2018)}]{Ji18}
{Ji}, A.~P., \& {Frebel}, A. 2018, \apj, 856, 138

\bibitem[{{Ji} {et~al.}(2016){Ji}, {Frebel}, {Chiti}, \& {Simon}}]{Ji16}
{Ji}, A.~P., {Frebel}, A., {Chiti}, A., \& {Simon}, J.~D. 2016, \nat, 531, 610

\bibitem[{Jones {et~al.}(2001)Jones, Oliphant, Peterson, {et~al.}}]{Jones01}
Jones, E., Oliphant, T., Peterson, P., {et~al.} 2001, {SciPy}: Open source
  scientific tools for {Python}, , , [Online; accessed 2017-01-16]

\bibitem[{{Karakas} \& {Lattanzio}(2014)}]{Karakas14}
{Karakas}, A.~I., \& {Lattanzio}, J.~C. 2014, PASA, 31, e030

\bibitem[{{Karakas} {et~al.}(2018){Karakas}, {Lugaro}, {Carlos}, {Cseh},
  {Kamath}, \& {Garc{\'{\i}}a-Hern{\'a}ndez}}]{Karakas18}
{Karakas}, A.~I., {Lugaro}, M., {Carlos}, M., {et~al.} 2018, \mnras, 477, 421

\bibitem[{{Kasen} \& {Bildsten}(2010)}]{Kasen10}
{Kasen}, D., \& {Bildsten}, L. 2010, \apj, 717, 245

\bibitem[{{Kirby} {et~al.}(2014){Kirby}, {Bullock}, {Boylan-Kolchin},
  {Kaplinghat}, \& {Cohen}}]{Kirby14}
{Kirby}, E.~N., {Bullock}, J.~S., {Boylan-Kolchin}, M., {Kaplinghat}, M., \&
  {Cohen}, J.~G. 2014, \mnras, 439, 1015

\bibitem[{{Kirby} {et~al.}(2011{\natexlab{a}}){Kirby}, {Cohen}, {Smith},
  {Majewski}, {Sohn}, \& {Guhathakurta}}]{Kirby11}
{Kirby}, E.~N., {Cohen}, J.~G., {Smith}, G.~H., {et~al.} 2011{\natexlab{a}},
  \apj, 727, 79

\bibitem[{{Kirby} {et~al.}(2009){Kirby}, {Guhathakurta}, {Bolte}, {Sneden}, \&
  {Geha}}]{Kirby09}
{Kirby}, E.~N., {Guhathakurta}, P., {Bolte}, M., {Sneden}, C., \& {Geha}, M.~C.
  2009, \apj, 705, 328

\bibitem[{{Kirby} {et~al.}(2016){Kirby}, {Guhathakurta}, {Zhang}, {Hong},
  {Guo}, {Guo}, {Cohen}, \& {Cunha}}]{Kirby16}
{Kirby}, E.~N., {Guhathakurta}, P., {Zhang}, A.~J., {et~al.} 2016, \apj, 819,
  135

\bibitem[{{Kirby} {et~al.}(2011{\natexlab{b}}){Kirby}, {Lanfranchi}, {Simon},
  {Cohen}, \& {Guhathakurta}}]{Kirby11b}
{Kirby}, E.~N., {Lanfranchi}, G.~A., {Simon}, J.~D., {Cohen}, J.~G., \&
  {Guhathakurta}, P. 2011{\natexlab{b}}, \apj, 727, 78

\bibitem[{{Kirby} {et~al.}(2010){Kirby}, {Guhathakurta}, {Simon}, {Geha},
  {Rockosi}, {Sneden}, {Cohen}, {Sohn}, {Majewski}, \& {Siegel}}]{Kirby10}
{Kirby}, E.~N., {Guhathakurta}, P., {Simon}, J.~D., {et~al.} 2010, \apjs, 191,
  352

\bibitem[{{Kirby} {et~al.}(2015){Kirby}, {Guo}, {Zhang}, {Deng}, {Cohen},
  {Guhathakurta}, {Shetrone}, {Lee}, \& {Rizzi}}]{Kirby15}
{Kirby}, E.~N., {Guo}, M., {Zhang}, A.~J., {et~al.} 2015, \apj, 801, 125

\bibitem[{{Kurucz}(1993)}]{Kurucz93}
{Kurucz}, R. 1993, ATLAS9 Stellar Atmosphere Programs and 2 km/s grid.~Kurucz
  CD-ROM No.~13.~ Cambridge, Mass.: Smithsonian Astrophysical Observatory,
  1993., 13

\bibitem[{{Lattimer} \& {Schramm}(1974)}]{Lattimer74}
{Lattimer}, J.~M., \& {Schramm}, D.~N. 1974, \apjl, 192, L145

\bibitem[{{Lee} {et~al.}(2005){Lee}, {Carney}, \& {Habgood}}]{Lee05}
{Lee}, J.-W., {Carney}, B.~W., \& {Habgood}, M.~J. 2005, \aj, 129, 251

\bibitem[{{Lee} {et~al.}(2003){Lee}, {Park}, {Park}, {Sohn}, {Oh}, {Yuk},
  {Rey}, {Lee}, {Lee}, {Kim}, {Han}, {Park}, {Lee}, {Jeon}, \& {Kim}}]{Lee03}
{Lee}, M.~G., {Park}, H.~S., {Park}, J.-H., {et~al.} 2003, \aj, 126, 2840

\bibitem[{{Lemasle} {et~al.}(2014){Lemasle}, {de Boer}, {Hill}, {Tolstoy},
  {Irwin}, {Jablonka}, {Venn}, {Battaglia}, {Starkenburg}, {Shetrone},
  {Letarte}, {Fran{\c c}ois}, {Helmi}, {Primas}, {Kaufer}, \&
  {Szeifert}}]{Lemasle14}
{Lemasle}, B., {de Boer}, T.~J.~L., {Hill}, V., {et~al.} 2014, \aap, 572, A88

\bibitem[{{Letarte} {et~al.}(2010){Letarte}, {Hill}, {Tolstoy}, {Jablonka},
  {Shetrone}, {Venn}, {Spite}, {Irwin}, {Battaglia}, {Helmi}, {Primas},
  {Fran{\c c}ois}, {Kaufer}, {Szeifert}, {Arimoto}, \& {Sadakane}}]{Letarte10}
{Letarte}, B., {Hill}, V., {Tolstoy}, E., {et~al.} 2010, \aap, 523, A17

\bibitem[{{Letarte, B.} {et~al.}(2018){Letarte, B.}, {Hill, V.}, {Tolstoy, E.},
  {Jablonka, P.}, {Shetrone, M.}, {Venn, K. A.}, {Spite, M.}, {Irwin, M. J.},
  {Battaglia, G.}, {Helmi, A.}, {Primas, F.}, {Fran\c{c}ois, P.}, {Kaufer, A.},
  {Szeifert, T.}, {Arimoto, N.}, \& {Sadakane, K.}}]{Letarte18}
{Letarte, B.}, {Hill, V.}, {Tolstoy, E.}, {et~al.} 2018, A\&A, 613, C1

\bibitem[{{Liccardo} {et~al.}(2018){Liccardo}, {Malheiro}, {Hussein}, \&
  {Frederico}}]{Liccardo18}
{Liccardo}, V., {Malheiro}, M., {Hussein}, M.~S., \& {Frederico}, T. 2018,
  ArXiv e-prints, arXiv:1805.10183

\bibitem[{{Macias} \& {Ramirez-Ruiz}(2016)}]{Macias16}
{Macias}, P., \& {Ramirez-Ruiz}, E. 2016, ArXiv e-prints, arXiv:1609.04826

\bibitem[{{Maoz} \& {Graur}(2017)}]{Maoz17}
{Maoz}, D., \& {Graur}, O. 2017, \apj, 848, 25

\bibitem[{{Mateo}(1998)}]{Mateo98}
{Mateo}, M.~L. 1998, \araa, 36, 435

\bibitem[{{McWilliam}(1998)}]{McWilliam98}
{McWilliam}, A. 1998, \aj, 115, 1640

\bibitem[{{Mighell} \& {Burke}(1999)}]{Mighell99}
{Mighell}, K.~J., \& {Burke}, C.~J. 1999, \aj, 118, 366

\bibitem[{{M{\"o}sta} {et~al.}(2017){M{\"o}sta}, {Roberts}, {Halevi}, {Ott},
  {Lippuner}, {Haas}, \& {Schnetter}}]{Mosta17}
{M{\"o}sta}, P., {Roberts}, L.~F., {Halevi}, G., {et~al.} 2017, ArXiv e-prints,
  arXiv:1712.09370

\bibitem[{{Newman} {et~al.}(2013){Newman}, {Cooper}, {Davis}, {Faber}, {Coil},
  {Guhathakurta}, {Koo}, {Phillips}, {Conroy}, {Dutton}, {Finkbeiner}, {Gerke},
  {Rosario}, {Weiner}, {Willmer}, {Yan}, {Harker}, {Kassin}, {Konidaris},
  {Lai}, {Madgwick}, {Noeske}, {Wirth}, {Connolly}, {Kaiser}, {Kirby},
  {Lemaux}, {Lin}, {Lotz}, {Luppino}, {Marinoni}, {Matthews}, {Metevier}, \&
  {Schiavon}}]{Newman13}
{Newman}, J.~A., {Cooper}, M.~C., {Davis}, M., {et~al.} 2013, \apjs, 208, 5

\bibitem[{{Nishimura} {et~al.}(2017){Nishimura}, {Sawai}, {Takiwaki}, {Yamada},
  \& {Thielemann}}]{Nishimura17}
{Nishimura}, N., {Sawai}, H., {Takiwaki}, T., {Yamada}, S., \& {Thielemann},
  F.-K. 2017, \apjl, 836, L21

\bibitem[{{Nishimura} {et~al.}(2015){Nishimura}, {Takiwaki}, \&
  {Thielemann}}]{Nishimura15}
{Nishimura}, N., {Takiwaki}, T., \& {Thielemann}, F.-K. 2015, \apj, 810, 109

\bibitem[{{Nomoto} {et~al.}(2006){Nomoto}, {Tominaga}, {Umeda}, {Kobayashi}, \&
  {Maeda}}]{Nomoto06}
{Nomoto}, K., {Tominaga}, N., {Umeda}, H., {Kobayashi}, C., \& {Maeda}, K.
  2006, Nuclear Physics A, 777, 424

\bibitem[{{Orban} {et~al.}(2008){Orban}, {Gnedin}, {Weisz}, {Skillman},
  {Dolphin}, \& {Holtzman}}]{Orban08}
{Orban}, C., {Gnedin}, O.~Y., {Weisz}, D.~R., {et~al.} 2008, \apj, 686, 1030

\bibitem[{{Papish} {et~al.}(2015){Papish}, {Soker}, \& {Bukay}}]{Papish15}
{Papish}, O., {Soker}, N., \& {Bukay}, I. 2015, \mnras, 449, 288

\bibitem[{{Pietrzy{\'n}ski} {et~al.}(2008){Pietrzy{\'n}ski}, {Gieren},
  {Szewczyk}, {Walker}, {Rizzi}, {Bresolin}, {Kudritzki}, {Nalewajko}, {Storm},
  {Dall'Ora}, \& {Ivanov}}]{Pietrzynski08}
{Pietrzy{\'n}ski}, G., {Gieren}, W., {Szewczyk}, O., {et~al.} 2008, \aj, 135,
  1993

\bibitem[{{Radice} {et~al.}(2016){Radice}, {Galeazzi}, {Lippuner}, {Roberts},
  {Ott}, \& {Rezzolla}}]{Radice16}
{Radice}, D., {Galeazzi}, F., {Lippuner}, J., {et~al.} 2016, \mnras, 460, 3255

\bibitem[{{Ram{\'{\i}}rez} \& {Allende Prieto}(2011)}]{Ramirez11}
{Ram{\'{\i}}rez}, I., \& {Allende Prieto}, C. 2011, \apj, 743, 135

\bibitem[{{Rizzi} {et~al.}(2007){Rizzi}, {Held}, {Saviane}, {Tully}, \&
  {Gullieuszik}}]{Rizzi07}
{Rizzi}, L., {Held}, E.~V., {Saviane}, I., {Tully}, R.~B., \& {Gullieuszik}, M.
  2007, \mnras, 380, 1255

\bibitem[{{Roederer} \& {Sneden}(2011)}]{RoedererSneden11}
{Roederer}, I.~U., \& {Sneden}, C. 2011, \aj, 142, 22

\bibitem[{{Ryabchikova} {et~al.}(2015){Ryabchikova}, {Piskunov}, {Kurucz},
  {Stempels}, {Heiter}, {Pakhomov}, \& {Barklem}}]{Ryabchikova15}
{Ryabchikova}, T., {Piskunov}, N., {Kurucz}, R.~L., {et~al.} 2015, \physscr,
  90, 054005

\bibitem[{{Sadakane} {et~al.}(2004){Sadakane}, {Arimoto}, {Ikuta}, {Aoki},
  {Jablonka}, \& {Tajitsu}}]{Sadakane04}
{Sadakane}, K., {Arimoto}, N., {Ikuta}, C., {et~al.} 2004, \pasj, 56, 1041

\bibitem[{{Safarzadeh} \& {Scannapieco}(2017)}]{Safarzadeh17}
{Safarzadeh}, M., \& {Scannapieco}, E. 2017, \mnras, 471, 2088

\bibitem[{{Shetrone} {et~al.}(2003){Shetrone}, {Venn}, {Tolstoy}, {Primas},
  {Hill}, \& {Kaufer}}]{Shetrone03}
{Shetrone}, M., {Venn}, K.~A., {Tolstoy}, E., {et~al.} 2003, \aj, 125, 684

\bibitem[{{Shetrone} {et~al.}(2001){Shetrone}, {C{\^o}t{\'e}}, \&
  {Sargent}}]{Shetrone01}
{Shetrone}, M.~D., {C{\^o}t{\'e}}, P., \& {Sargent}, W.~L.~W. 2001, \apj, 548,
  592

\bibitem[{{Shibata} \& {Taniguchi}(2011)}]{Shibata11}
{Shibata}, M., \& {Taniguchi}, K. 2011, Living Reviews in Relativity, 14, 6

\bibitem[{{Simmerer} {et~al.}(2004){Simmerer}, {Sneden}, {Cowan}, {Collier},
  {Woolf}, \& {Lawler}}]{Simmerer04}
{Simmerer}, J., {Sneden}, C., {Cowan}, J.~J., {et~al.} 2004, \apj, 617, 1091

\bibitem[{{Simon} {et~al.}(2015){Simon}, {Jacobson}, {Frebel}, {Thompson},
  {Adams}, \& {Shectman}}]{Simon15}
{Simon}, J.~D., {Jacobson}, H.~R., {Frebel}, A., {et~al.} 2015, \apj, 802, 93

\bibitem[{{Sk{\'u}lad{\'o}ttir} {et~al.}(2015){Sk{\'u}lad{\'o}ttir}, {Tolstoy},
  {Salvadori}, {Hill}, {Pettini}, {Shetrone}, \& {Starkenburg}}]{Skuladottir15}
{Sk{\'u}lad{\'o}ttir}, {\'A}., {Tolstoy}, E., {Salvadori}, S., {et~al.} 2015,
  \aap, 574, A129

\bibitem[{{Sneden} {et~al.}(2012){Sneden}, {Bean}, {Ivans}, {Lucatello}, \&
  {Sobeck}}]{Sneden12}
{Sneden}, C., {Bean}, J., {Ivans}, I., {Lucatello}, S., \& {Sobeck}, J. 2012,
  {MOOG: LTE line analysis and spectrum synthesis}, Astrophysics Source Code
  Library, , , ascl:1202.009

\bibitem[{{Sneden} {et~al.}(2008){Sneden}, {Cowan}, \& {Gallino}}]{Sneden08}
{Sneden}, C., {Cowan}, J.~J., \& {Gallino}, R. 2008, \araa, 46, 241

\bibitem[{{Sneden} {et~al.}(1992){Sneden}, {Kraft}, {Prosser}, \&
  {Langer}}]{Sneden92}
{Sneden}, C., {Kraft}, R.~P., {Prosser}, C.~F., \& {Langer}, G.~E. 1992, \aj,
  104, 2121

\bibitem[{{Sneden} {et~al.}(1997){Sneden}, {Kraft}, {Shetrone}, {Smith},
  {Langer}, \& {Prosser}}]{Sneden97}
{Sneden}, C., {Kraft}, R.~P., {Shetrone}, M.~D., {et~al.} 1997, \aj, 114, 1964

\bibitem[{{Sneden} {et~al.}(2014){Sneden}, {Lucatello}, {Ram}, {Brooke}, \&
  {Bernath}}]{Sneden14}
{Sneden}, C., {Lucatello}, S., {Ram}, R.~S., {Brooke}, J.~S.~A., \& {Bernath},
  P. 2014, \apjs, 214, 26

\bibitem[{{Sneden}(1973)}]{Sneden73}
{Sneden}, C.~A. 1973, PhD thesis, THE UNIVERSITY OF TEXAS AT AUSTIN.

\bibitem[{{Soker} \& {Gilkis}(2017)}]{Soker17}
{Soker}, N., \& {Gilkis}, A. 2017, \apj, 851, 95

\bibitem[{{Spite}(1967)}]{Spite67}
{Spite}, M. 1967, Annales d'Astrophysique, 30, 211

\bibitem[{{Starkenburg} {et~al.}(2013){Starkenburg}, {Hill}, {Tolstoy},
  {Fran{\c c}ois}, {Irwin}, {Boschman}, {Venn}, {de Boer}, {Lemasle},
  {Jablonka}, {Battaglia}, {Groot}, \& {Kaper}}]{Starkenburg13}
{Starkenburg}, E., {Hill}, V., {Tolstoy}, E., {et~al.} 2013, \aap, 549, A88

\bibitem[{{Suda} {et~al.}(2017){Suda}, {Hidaka}, {Aoki}, {Katsuta}, {Yamada},
  {Fujimoto}, {Ohtani}, {Masuyama}, {Noda}, \& {Wada}}]{Suda17}
{Suda}, T., {Hidaka}, J., {Aoki}, W., {et~al.} 2017, \pasj, 69, 76

\bibitem[{{Tafelmeyer} {et~al.}(2010){Tafelmeyer}, {Jablonka}, {Hill},
  {Shetrone}, {Tolstoy}, {Irwin}, {Battaglia}, {Helmi}, {Starkenburg}, {Venn},
  {Abel}, {Francois}, {Kaufer}, {North}, {Primas}, \&
  {Szeifert}}]{Tafelmeyer10}
{Tafelmeyer}, M., {Jablonka}, P., {Hill}, V., {et~al.} 2010, \aap, 524, A58

\bibitem[{{Takeda}(1995)}]{Takeda95}
{Takeda}, Y. 1995, \pasj, 47, 287

\bibitem[{{Tanvir} {et~al.}(2017){Tanvir}, {Levan},
  {Gonz{\'a}lez-Fern{\'a}ndez}, {Korobkin}, {Mandel}, {Rosswog}, {Hjorth},
  {D'Avanzo}, {Fruchter}, {Fryer}, {Kangas}, {Milvang-Jensen}, {Rosetti},
  {Steeghs}, {Wollaeger}, {Cano}, {Copperwheat}, {Covino}, {D'Elia}, {de Ugarte
  Postigo}, {Evans}, {Even}, {Fairhurst}, {Figuera Jaimes}, {Fontes}, {Fujii},
  {Fynbo}, {Gompertz}, {Greiner}, {Hodosan}, {Irwin}, {Jakobsson},
  {J{\o}rgensen}, {Kann}, {Lyman}, {Malesani}, {McMahon}, {Melandri},
  {O'Brien}, {Osborne}, {Palazzi}, {Perley}, {Pian}, {Piranomonte}, {Rabus},
  {Rol}, {Rowlinson}, {Schulze}, {Sutton}, {Th{\"o}ne}, {Ulaczyk}, {Watson},
  {Wiersema}, \& {Wijers}}]{Tanvir17}
{Tanvir}, N.~R., {Levan}, A.~J., {Gonz{\'a}lez-Fern{\'a}ndez}, C., {et~al.}
  2017, \apjl, 848, L27

\bibitem[{{Thielemann} {et~al.}(2011){Thielemann}, {Arcones}, {K{\"a}ppeli},
  {Liebend{\"o}rfer}, {Rauscher}, {Winteler}, {Fr{\"o}hlich}, {Dillmann},
  {Fischer}, {Martinez-Pinedo}, {Langanke}, {Farouqi}, {Kratz}, {Panov}, \&
  {Korneev}}]{Thielemann11}
{Thielemann}, F.-K., {Arcones}, A., {K{\"a}ppeli}, R., {et~al.} 2011, Progress
  in Particle and Nuclear Physics, 66, 346

\bibitem[{{Tinsley}(1980)}]{Tinsley80}
{Tinsley}, B.~M. 1980, \fcp, 5, 287

\bibitem[{{Tolstoy} {et~al.}(2009){Tolstoy}, {Hill}, \& {Tosi}}]{Tolstoy09}
{Tolstoy}, E., {Hill}, V., \& {Tosi}, M. 2009, \araa, 47, 371

\bibitem[{{Tolstoy} {et~al.}(2006){Tolstoy}, {Hill}, {Irwin}, {Helmi},
  {Battaglia}, {Letarte}, {Venn}, {Jablonka}, {Shetrone}, {Arimoto}, {Abel},
  {Primas}, {Kaufer}, {Szeifert}, {Francois}, \& {Sadakane}}]{Tolstoy06}
{Tolstoy}, E., {Hill}, V., {Irwin}, M., {et~al.} 2006, The Messenger, 123

\bibitem[{{Troja} {et~al.}(2017){Troja}, {Piro}, {van Eerten}, {Wollaeger},
  {Im}, {Fox}, {Butler}, {Cenko}, {Sakamoto}, {Fryer}, {Ricci}, {Lien}, {Ryan},
  {Korobkin}, {Lee}, {Burgess}, {Lee}, {Watson}, {Choi}, {Covino}, {D'Avanzo},
  {Fontes}, {Gonz{\'a}lez}, {Khandrika}, {Kim}, {Kim}, {Lee}, {Lee}, {Kutyrev},
  {Lim}, {S{\'a}nchez-Ram{\'{\i}}rez}, {Veilleux}, {Wieringa}, \&
  {Yoon}}]{Troja17}
{Troja}, E., {Piro}, L., {van Eerten}, H., {et~al.} 2017, \nat, 551, 71

\bibitem[{{Tsuji}(1978)}]{Tsuji78}
{Tsuji}, T. 1978, \aap, 62, 29

\bibitem[{{Tsujimoto} {et~al.}(2017){Tsujimoto}, {Matsuno}, {Aoki}, {Ishigaki},
  \& {Shigeyama}}]{Tsujimoto17}
{Tsujimoto}, T., {Matsuno}, T., {Aoki}, W., {Ishigaki}, M.~N., \& {Shigeyama},
  T. 2017, \apjl, 850, L12

\bibitem[{{Tsujimoto} \& {Nishimura}(2015)}]{Tsujimoto15_mrsn}
{Tsujimoto}, T., \& {Nishimura}, N. 2015, \apjl, 811, L10

\bibitem[{{Ural} {et~al.}(2015){Ural}, {Cescutti}, {Koch}, {Kleyna},
  {Feltzing}, \& {Wilkinson}}]{Ural15}
{Ural}, U., {Cescutti}, G., {Koch}, A., {et~al.} 2015, \mnras, 449, 761

\bibitem[{{Venn} {et~al.}(2004){Venn}, {Irwin}, {Shetrone}, {Tout}, {Hill}, \&
  {Tolstoy}}]{Venn04}
{Venn}, K.~A., {Irwin}, M., {Shetrone}, M.~D., {et~al.} 2004, \aj, 128, 1177

\bibitem[{{Venn} {et~al.}(2012){Venn}, {Shetrone}, {Irwin}, {Hill}, {Jablonka},
  {Tolstoy}, {Lemasle}, {Divell}, {Starkenburg}, {Letarte}, {Baldner},
  {Battaglia}, {Helmi}, {Kaufer}, \& {Primas}}]{Venn12}
{Venn}, K.~A., {Shetrone}, M.~D., {Irwin}, M.~J., {et~al.} 2012, \apj, 751, 102

\bibitem[{{Wanajo}(2013)}]{Wanajo13}
{Wanajo}, S. 2013, \apjl, 770, L22

\bibitem[{{Wehmeyer} {et~al.}(2015){Wehmeyer}, {Pignatari}, \&
  {Thielemann}}]{Wehmeyer15}
{Wehmeyer}, B., {Pignatari}, M., \& {Thielemann}, F.-K. 2015, \mnras, 452, 1970

\bibitem[{{Weisz} {et~al.}(2014){Weisz}, {Dolphin}, {Skillman}, {Holtzman},
  {Gilbert}, {Dalcanton}, \& {Williams}}]{Weisz14}
{Weisz}, D.~R., {Dolphin}, A.~E., {Skillman}, E.~D., {et~al.} 2014, \apj, 789,
  148

\bibitem[{{Willems} \& {Kalogera}(2004)}]{Willems04}
{Willems}, B., \& {Kalogera}, V. 2004, \apjl, 603, L101

\bibitem[{{Woo} {et~al.}(2008){Woo}, {Courteau}, \& {Dekel}}]{Woo08}
{Woo}, J., {Courteau}, S., \& {Dekel}, A. 2008, \mnras, 390, 1453

\bibitem[{{Woosley} \& {Janka}(2005)}]{Woosley05}
{Woosley}, S., \& {Janka}, T. 2005, Nature Physics, 1, 147

\bibitem[{{Worley} {et~al.}(2013){Worley}, {Hill}, {Sobeck}, \&
  {Carretta}}]{Worley13}
{Worley}, C.~C., {Hill}, V., {Sobeck}, J., \& {Carretta}, E. 2013, \aap, 553,
  A47

\end{thebibliography}

\listofchanges

\end{document}